\documentclass[fleqn,usenatbib]{mnras}
\usepackage[utf8]{inputenc}

\usepackage{newtxtext,newtxmath}

\usepackage[T1]{fontenc}
\usepackage{comment}
\usepackage{siunitx}

\DeclareRobustCommand{\VAN}[3]{#2}
\let\VANthebibliography\thebibliography
\def\thebibliography{\DeclareRobustCommand{\VAN}[3]{##3}\VANthebibliography}

\usepackage{graphicx}	
\usepackage{amsmath}	
\usepackage{orcidlink}

\title[Multi-planet detections with Gaia]{Detecting multiple planets in Gaia data using Bayesian model selection}

\author[Greenfield \& Owen]{
Emmanuel J. Greenfield\orcidlink{0009-0002-4021-062X}\thanks{E-mail: eg319@ic.ac.uk}
and James E. Owen\orcidlink{0000-0002-4856-7837}
\\
Imperial Astrophysics, Department of Physics, Imperial College London, Prince Consort Road, London SW7 2AZ, UK\\
}

\date{Accepted XXX. Received YYY; in original form ZZZ}

\pubyear{2026}

\begin{document}
\newcommand{\rc}[1]{\textcolor{purple}{#1}}

\label{firstpage}
\pagerange{\pageref{firstpage}--\pageref{lastpage}}
\maketitle

\begin{abstract}

Gaia will soon provide a census of cold giant planets. However, an astrometric signal contains the superimposed reflex motion of all companions, so planetary signals cannot be treated individually. We therefore investigate Gaia's planet-detection capabilities for 1-planet and 2-planet systems, with special attention to systems near 2:1 mean-motion resonance. Gaia can detect 2-planet systems, although the lower-S/N planet requires a higher S/N threshold, especially when the outer planet is more massive and near 2:1 commensurability. When fitting a 1-planet model to a circular 2-planet resonant system, the resulting fit shows biases in eccentricity (up to e = 0.8), mass (up to 100\%), and inclination (up to 50 deg). The bias is maximised when the planets' astrometric signatures are similar, since the combined orbital signal deviates most from a single Keplerian orbit. Many of these systems can be identified as having 2 planets using Bayesian model selection. Still, in some cases the second planet is hidden and undetectable due to its S/N, leading to these biases for the outer planet's orbital parameters. When the outer planet is roughly over three times as massive as the inner planet, orbital approximations allow us to recover the astrometric resoeccentric degeneracy from the bias. When the planets are in dynamical resonance, apsidal precession observed over a long enough baseline (e.g. DR5) can contribute to breaking the degeneracy at short orbital periods. Overall, these biases might significantly affect constraints on planet formation, migration, and evolution models from Gaia population-wide demographics.

\end{abstract}




\section{Introduction}

Since the discovery of the first giant exoplanet \citep[][]{mayor_jupiter-mass_1995}, the known landscape of exoplanet populations has expanded and diversified considerably. The now thousands of confirmed exoplanet systems have been pivotal in informing planet formation, migration and evolution models. Specifically, the Kepler \citep[][]{borucki_kepler_2010, thompson_planetary_2018} and now TESS \citep[][]{ricker_transiting_2015,guerrero_tess_2021} missions have provided valuable information on transiting planets close to their host star. Longer period planets, however, are much harder to detect through the transit method, due to their low transit probability and long orbital periods. It is hence harder to study these systems on the level of population demographics. Core accretion giant planet formation models \citep[e.g.][]{pollack_formation_1996} predict that giant planets form at distances of around 10 au and migrate inwards. However, we still struggle to constrain the exact mechanisms that limit the planet mass once the phase of runaway gas accretion is reached \citep[e.g.][]{mizuno_formation_1980, rafikov_atmospheres_2006, bailey_growing_2024}. Consequently, the current observed semi-major axis and eccentricity distributions are difficult to explain with formation and evolution models \citep[e.g.][]{jennings_comparative_2018, bitsch_eccentricity_2020}. Observations are thus essential to better inform these models. We also suspect that distant gas giants, analogous to Jupiter and Saturn in mass and semi-major axis, have had a major influence in shaping the form of the early solar system planets \citep[e.g.][]{tsiganis_origin_2005,nesvorny_statistical_2012,lau_sequential_2024}, further motivating the need to study giant exoplanets in solar system analogues.

Various detection techniques cover complementary regions of mass-period parameter space. For example, microlensing covers the large orbital separation detection space. Only a limited number of systems ($\sim$ 200) have been observed so far \citep{mroz_exoplanet_2024}, and the demographics from certain of these observations indicate a negative power law \citep[][]{cassan_one_2012} or a broken power law \citep[][]{suzuki_exoplanet_2016} distribution in planet-star mass ratio for microlensed planets, with a particularly large occurrence of super-Earths on Jupiter-like orbits \citep[][]{zang_microlensing_2025}. It is much harder to establish the same for semi-major axis, given that microlensing is only capable of measuring a projected separation.

Currently, our best knowledge of cold giant planet demographics comes from radial velocity (RV) surveys such as the California Legacy Survey \citep[][]{fulton_california_2021}, and the \citet{mayor_harps_2011} survey analysed by \citet{fernandes_hints_2019}. These surveys find an increase in giant planet occurrence rate up to a few au, after which they find a tentative decline, with a limited amount of data beyond 5 au. Planets beyond this separation are harder to find due to the longer baselines required to accurately constrain their properties.  In agreement with the aforementioned microlensing results, the mass function is found in \citet{fernandes_hints_2019} to be a power law, although with a less steep gradient. Current RV samples estimate the mean eccentricity of such cold giant planets beyond the ice line to be around 0.23 \citep{kane_eccentricity_2024,alqasim_investigating_2025}. The origin of this distribution is unknown, as it can be a consequence of formation \citep[][]{bitsch_eccentricity_2020} or come from mischaracterising systems in 2:1 mean motion resonance (MMR) as a single eccentric Keplerian \citep[][]{anglada-escude_how_2010}.

The Gaia mission \citep[][]{bernstein_finding_1995,casertano_astrometric_1995, gaia_collaboration_gaia_2016} will provide a step-change in the exploration for the detection-space of cold giant planets. With a final observation baseline of over 10 years, it will accurately constrain the masses and orbital architectures of planets with semi-major axes up to $\sim$10 au, where they are currently thought to form. Predictive studies such as \citet{perryman_astrometric_2014} and \citet{lammers_exoplanet_2025} suggest that Gaia will discover 10 000 to 100 000 planets by the end of the current decade. Astrometric time series of each star are currently unavailable, pending Data Release 4 (DR4) in December 2026, but results from DR3 show that Gaia should reach its predicted precision of a few tens of $\mu$ as. To prepare for the release of DR4, many studies have considered the detectability criteria for a single planet \citep[e.g.][]{ranalli_astrometry_2018}, but the case of multiple planets is often omitted. 

Giant planets are not predicted to form in isolation, and young systems such as PDS 70 \citep[][]{haffert_two_2019} or HR 8799 \citep[][]{marois_images_2010} demonstrate this explicitly. Amongst the limited number of studies that investigate Gaia's multi-planet detection capabilities, the study by \citet{sozzetti_detection_2001} showed that Gaia can detect multiple planets in the specific system of $\upsilon$ Andromedæ. \citet{casertano_double-blind_2008} then generalised the argument and showed that Gaia can successfully detect most two-planet systems with well separated orbital periods, disregarding any resonances. The study by \citet{wu_detecting_2016} suggested limited recovery capabilities for systems near resonance, but greatly simplified the data recovery from Gaia.

Unlike transits, where contamination of the signal from other planets in the systems is limited and reasonably well characterised \citep[e.g.][]{Zink2019}, astrometry is a dynamical measurement of the evolution of the system photocentre and is hence affected by all planets at all times. On top of completing the giant planet population demographics, we need to improve our understanding of multi-planet system architectures. This need is motivated by the observation that many close-in systems are in resonance or near-resonant \citep[][]{fabrycky_architecture_2014, winn_occurrence_2015}, with a higher resonant fraction at earlier times \citep[e.g.][]{Dai2024}. Furthermore, resonances could be a natural outcome of giant planet formation and disc dispersal \citep[e.g.][]{Greenfield2026}. Planets close to resonance (i.e. near commensurability) will generate a highly correlated signal. This signal will lead to a fixed repeating photocentre pattern that may likely impact the detectability of multi-planet systems \citep[][]{yahalomi_astrometric_2025}. Similar degeneracies have been found in the RV, where signals are similarly correlated \citep[][]{anglada-escude_how_2010}.

Motivated by the need to uncover multi-planet systems in Gaia data, we develop a robust framework for astrometric planet detection, including a Bayesian model preference test to determine whether a 0, 1 or 2 planet model\footnote{In this paper we focus on systems of up to two planets as a first step; this framework is easily generalisable to more planets.} is the best fit to the available data. Studies like \citet{ranalli_astrometry_2018} and  \citet{holl_gaia_2023} have attempted to use the Bayesian Information Criterion (BIC) comparing a single planet model to a zero planet model, instead of the more commonly used $\Delta\chi^2$ test, but the BIC fails to fully represent the posterior shapes for each of the orbital parameters \citep[][]{liddle_information_2007}. Bayes factor model preference testing is hence essential if we want to limit the contamination of orbital parameters by other planets. Given DR4 will predominantly contain single planet fits, we use our framework to also investigate the impact of a 2 planet signal on a single planet fit. 

We have organised the paper as follows. In Section \ref{sec:methods} we explain detection criteria, how we generate mock Gaia signals, and how we set up the nested sampling detection pipeline. In Section \ref{sec:1_planet} we test the pipeline, confirm that it works for sample DR4 data, and recover previously found detection criteria. Section \ref{sec:multiple_planet_detections} focuses on multi-planet detections. We examine the 2-planet detection space near and far from 2:1 commensurability. We then fit 1- and 2-planet models to these signals and discuss the emergent orbital parameter biases and degeneracies that follow from the planets being near and in exact dynamical resonance. Section \ref{sec:discussion} is a short discussion on the wider context and implications of the previous findings. Finally, in Section \ref{sec:conclusions} we summarise our results and present our conclusions.

\section{A nested sampling pipeline for astrometric single planet detections} \label{sec:methods}

\subsection{Detection criteria}

Before moving to multi-planet systems, we first want to consider the astrometric detectability of a single planet. Astrometric detections of exoplanets rely on resolving the reflex motion of the star due to its companion. Given that both objects orbit the barycentre of the system, we can consider the astrometric signature of the star given by:

\begin{equation} \label{eq:signature}
    \left(\frac{\upsilon}{1~\text{arcsec}}\right) = \left(\frac{m_\text{p}}{m_{\star}}\right)\left(\frac{a_{\text{p}}}{ \text{1 au}}\right)\left(\frac{d}{\text{1 pc}}\right)^{-1},
\end{equation}

where $m_\text{p}$ is the mass of the planet, $m_{\star}$ is the mass of the star, $a_{\text{p}}$ is the planet's semi-major axis and $d$ is the distance of the system to Earth. This quantity corresponds to the projected semi-amplitude of the stellar displacement due to the planet, as seen from Earth. As an example, a planet like Jupiter orbiting the Sun at 50 pc would have $\upsilon_{\text{Jup}} = 100 \: \mu$as. Previous studies \citep[e.g.][]{casertano_double-blind_2008,perryman_astrometric_2014} have used the signal to noise in a single astrometric point, S/N, to determine detectability, where:

\begin{equation}
    S/N = \frac{\upsilon}{\sigma_{\text{AL}}}
\end{equation}

where $\sigma_{\text{AL}}$ is the Gaia along-scan uncertainty per field of view crossing. This error has random and systematic components, including a dependence on the star's apparent G-band magnitude (G). For G > 13, photon-counting noise becomes increasingly dominant \citep{lindegren_gaia_2021}, but for G < 12, the overall uncertainty remains constant, at around $\sim 30 \: \mu$as as predicted in \citet{perryman_astrometric_2014}, and later confirmed in \citet{holl_gaia_2023} from the Gaia DR3 data to be  $\sim 40 \: \mu$as. Anticipating further improvements to this for DR4 and DR5, we assume $\sigma_{\text{AL}} = 30 \: \mu$as for the remainder of this study. Initial studies \citep[e.g.][]{casertano_double-blind_2008} establish that planets are detectable with a S/N > 3, but subsequent research \citep[e.g.][]{perryman_astrometric_2014, ranalli_astrometry_2018, lammers_exoplanet_2025} shows that this criterion is much looser, with an absolute lower limit at S/N >1 for a 10 year mission (DR5). For $1<S/N<3$, they conclude that detections are possible but not guaranteed, and the mass cannot be resolved at a precision better than 20\%. Here we aim to confirm these results within a Bayesian model comparison framework, and to verify the accuracy of our pipeline.\\

The work is divided into two parts; first we generate mock Gaia observation data; second we fit the data to a single-planet and zero-planet model, including proper motion, parallax and celestial coordinates offset. We then do a model preference test using nested sampling.  

\subsection{Observing stars with Gaia}

A star without a companion will have a distinct astrometric path in the sky, and its celestial coordinates $(\alpha^* (t),\delta (t)) = (\alpha (t) \cos \delta_0, \delta (t))$ evolve as follows, from a reference point ($\alpha_0, \delta_0$) at time $t=t_0$
\begin{equation} \label{astrometryeq}
    \begin{aligned}
        &\alpha^* (t) = \alpha_0^* +\Delta\alpha_0^* + \mu_{\alpha^*} (t-t_0) +  \Pi_{\alpha^*} p\\
        &\delta (t)= \delta_0 +\Delta\delta_0 + \mu_\delta (t-t_0) +  \Pi_\delta p
    \end{aligned}
\end{equation}
This astrometric motion is described by 5 parameters: $\Delta\alpha_0^*$ and $\Delta\delta_0$ are the initial (small) displacements in right ascension and declination from the reference position, $\mu_{\alpha^*}$ and $\mu_\delta$ are the proper motion components in right ascension and declination, encoding a linear displacement in the sky with respect to the observer. Finally, $p=1/d$ is the parallax, encoding the distance $d$ of the system to the observer. $\Pi_{\alpha^*}$ and  $\Pi_\delta$ are the (known) parallax factors, representing the periodic motion of the observer around the Sun.

Gaia rotates and measures the position of stars as they cross the CCDs; hence the error in the along-scan direction is much smaller than in the across-scan direction, which is subject to heavy binning \citep[][]{lindegren_astrometric_2012}. We hence consider the measurement to be the projection of the celestial coordinates on a single line, defined by the precession of the satellite and given by the Gaia scanning law, similar to the Gaia astrometric fitting pipelines \citep[][]{lindegren_gaia_2021,halbwachs_gaia_2023}. The projected coordinate $\nu_i$ measured  at time $t_i$ is given by:
\begin{equation} \label{projectioneq}
    \nu_i = \alpha_i^*\sin{\theta_i}+\delta_i\cos{\theta_i}
\end{equation}
where $\theta_i$ is the along-scan scan angle. 

\subsection{Generating astrometric observations} \label{sec:generating1pl}

\begin{table}
	\centering
	\caption{Sample 5-parameter astrometry used to generate mock observations}
	\label{tab:mock_astrometry}
	\begin{tabular}{lcccr} 
		\hline
		  $\Delta\alpha_0^*$ (mas) & $\Delta\delta_0$ (mas) & $\mu_{\alpha^*}$ (mas/yr) & $\mu_\delta$ (mas/yr) & $d$ (pc)\\
		\hline
		 $2.02 \times 10^{-2}$ & $1.47 \times 10^{-2}$& 5 & -24 & 16 \\
		\hline
	\end{tabular}
\end{table}

Despite fitting Keplerian orbits later on, we choose to generate the mock planetary signals as N-body orbits as this will allow us to test the robustness of the approximation. This test is essential to do for planets near MMRs due to strong planet-planet interactions. We use the N-body code REBOUND with the \texttt{ias15} integrator \citep[][]{rein_rebound_2012,rein_ias15_2015}. We specify the initial semi-major axis, mass and eccentricity of the planet (in units of au and stellar mass respectively). To best simulate the realistic cadence of Gaia observations, we select a representative point $(\alpha,\delta) = (\pi/4,\pi/4)$ and use the online Gaia Observation Forecast Tool (GOST) \footnote{\url{https://gaia.esac.esa.int/gost/}} to extract observation times, scan angles and parallax factors for each of the 172 observations over the full mission lifetime. We confirm that this specific point is representative of the typical number of Gaia observations by running the calculations described in the next sections with a sample of different reference coordinates and find no major quantitative difference in the results. 
We then integrate the system until the given observation time (either DR4, or DR5) and record the position of the star in the centre-of-mass frame $(x_\text{com},y_\text{com}) $. For simplicity, we set up the systems face-on to maximise detectability.

We choose a representative 5-parameter astrometry for the star, as described in Table \ref{tab:mock_astrometry}, with a system located at 16 pc. Using Eq. \ref{astrometryeq}, we generate the corresponding astrometric signal for the Gaia observation times, to which we add the deviation $(\Delta \alpha_{\text{pl}}^*,\Delta \delta_{\text{pl}})=(y_{\text{com}}p, x_{\text{com}}p) $, due to the planetary signal that was previously generated.

The error is sampled from a one-dimensional Gaussian distribution, centred on 0 and with a standard deviation of $30 \mu$ as. This sampling is done independently for each observation. This resulting photocentre trajectory is then projected in celestial coordinates using the known scan angle, and then added to the previously generated signal.

The overall signal, including the sum of the stellar astrometry, reflex motion and generated error, is then projected back into the one-dimensional Gaia scan error direction using Eq. \ref{projectioneq}. 

\subsection{Setting up the detection pipeline}\label{sec:model1pl}

For a single planet, the N-body signal is a Keplerian orbit. For multiple planets, as discussed later on, there are deviations from Keplerian motion that we assume to be negligible for the purpose of detecting and characterising the planetary signals. The reflex motion due to the planet is hence described as a Keplerian orbit $\textbf{r}_{\text{2D}}(t) = (x(t),y(t),z(t))$ in the (x,y) orbital plane \citep[][]{murray_solar_1999}:
\begin{equation}
    \begin{aligned}
     &x(t)= a_{\star} \left(\cos E(t) -e\right)\\
     &y(t)=a_{\star} \sqrt{1-e^2} \sin E(t) \\
     &z(t)=0
    \end{aligned}
    \label{eq:kepl_orbit}
\end{equation}
where $a_{\star} = q/(1+q)\, a$ is the semi-major axis of the star (with $q = m_\text{p}/m_{\star}$ and $a$ is the relative planet-star semi-major axis, i.e. $a=a_\text{p}+a_\star$), $e$ is the eccentricity of the planet and $E(t)$ is the eccentric anomaly, given by the solution to Kepler's equation:
\begin{equation}
    M(t) = M_0 + \frac{2 \pi}{P} (t - \tau) = E(t) - e \sin E(t),
\end{equation}
where $M(t)$ and $M_0$ are the mean anomalies at times $t$ and $\tau$ respectively, $P$ is the orbital period and $\tau$ is a reference time. These equations contain 4 out of the 7 parameters used to describe an orbit. The three remaining parameters are the argument of the stellar pericentre ($\omega$), the inclination ($I$) and the longitude of the ascending node ($\Omega$). They are included in the orbit through a series of rotations:
\begin{equation}
    \textbf{r}(t) = (X(t),Y(t),Z(t)) = R_z(\Omega)R_x(I)R_z(\omega) \textbf{r}_{\text{2D}}(t)
\end{equation}
where \textbf{r}(t) describes the orbit in the sky plane, and $R_x$ and $R_z$ are the rotation matrices along the $x$ and $z$ directions respectively. We can then write:
\begin{equation} \label{eq:XY}
    \begin{aligned}
     &X(t)= A'x(t)+ F'y(t)\\
     &Y(t)=B'x(t)+ G'y(t)
    \end{aligned}
\end{equation}
where $A',B',F'$ and $G'$ are the reduced Thiele-Innes constants \citep[][]{feng_astrometric_2024}, encapsulating the rotations mentioned previously ($R_z(\Omega)R_x(I)R_z(\omega)$), and expressed as:
\begin{equation}
    \begin{aligned}
        A' &= \cos \Omega \cos \omega - \sin \Omega \sin \omega \cos I \\
        B' &= \sin \Omega \cos \omega + \cos \Omega \sin \omega \cos I \\
        F' &= -\cos \Omega \sin \omega - \sin \Omega \cos \omega \cos I \\
        G' &= -\sin \Omega \sin \omega + \cos \Omega \cos \omega \cos I
    \end{aligned}
\end{equation}
Hence, the full observed deviation in celestial coordinates due to a planet is given by: 
\begin{equation} \label{eq:ra_decl}
    (\Delta\alpha_{\text{pl}}^*,\Delta\delta_{\text{pl}}) = (Y(t)p,X(t)p).
\end{equation}

\subsection{Nested sampling}

We use the \texttt{dynesty} nested sampling package \citep[][]{speagle_dynesty_2020} to fit the mock data generated in Section \ref{sec:generating1pl} to the model presented in Section \ref{sec:model1pl}. We adopt the following form for the likelihood

\begin{equation} \label{eq:likelihood}
    \log(\mathcal{L}) = - \sum_i^N\log(\sqrt{2\pi} \sigma_{i}) -\sum_i^N \frac{(\nu_{\text{obs}_{i}}-\nu_{\text{model}_{i}})^2}{2\sigma_{i}^2}, 
\end{equation}
where N is the number of observations, $\nu_{\text{obs}_{i}}$ is the along-scan coordinate of the observation, $\nu_{\text{model}_{i}}$ is the along-scan coordinate given by the model and $\sigma_{i}$ is the uncertainty corresponding to each observation (set to $\sigma_{\text{AL}}$ for the mock generated data). We build a fitting routine for two different models: one for a star without a planet (5 parameters) and one for a single planet around a star (12 parameters). We extend these models to multiple planets in Section \ref{sec:multiple_planet_detections}. The priors for each parameter are indicated in Table \ref{tab:priors}. We have found that recasting $\Delta \alpha_0^* = \Delta r \cos \Delta \theta$ and  $\Delta \delta_0 = \Delta r \sin \Delta \theta$ allows us to better implement a log-uniform prior radially around ($\alpha_0^*$,$\delta_0$). We adjust our upper and lower prior bounds for mass and semi-major axis based on the current known detection limits for Gaia. We use 2500 live points, which we found is enough to consistently recover most systems. For applications to real data, one must carefully evaluate this threshold, as it is only good for a reliable parameter space study, but we found it can still fail at the level of some individual systems.

\begin{table}
	\centering
	\caption{List of priors for each parameter in the simple planet retrieval model}
	\label{tab:priors}
	\begin{tabular}{cc} 
		\hline
		  Parameter & Distribution\\
		\hline
	$\Delta r$ (arcsec) & Log-Uniform($10^{-6}$, 1) \\
    $\Delta \theta$ & Uniform(0, 2$\pi$) \\
    $\mu_{\alpha^*}$ (arcsec/yr) & Uniform(-1, 1) \\
    $\mu_\delta$ (arcsec/yr) & Uniform(-1, 1) \\
    $p$ (arcsec) & Log-Uniform($10^{-4}$, 1) \\
    \hline
    $a$ (au) & Log-Uniform(0.05, 200) \\
    $\omega$ & Uniform(0, 2$\pi$) \\
    $\Omega$ & Uniform(0, $\pi$) \\
    $\cos I$ & Uniform(-1, 1) \\
    $M_0$ & Uniform(0, 2$\pi$) \\
    $e$ & Uniform(0, 0.9999) \\
    $q$ & Log-Uniform($5\times10^{-5}$, 0.1) \\
		\hline
	\end{tabular}
\end{table}

Previous studies have used Markov Chain Monte Carlo (MCMC) methods to fit the astrometric and orbital parameters of the planet \citep[e.g.][]{lammers_exoplanet_2025}. This also applies to the Gaia DR4 fitting pipeline, which uses a periodogram and MCMC joint approach to characterising exoplanets. This method has advantages and disadvantages. Recasting orbital parameters in Thiele-Innes constants creates a linear problem to solve in the orbital angles, hence making MCMC fitting routines much faster. Nested sampling, however, has its own significant advantages. 

Firstly, it is more accurate when it comes to model selection. Most previous studies adopt a frequentist approach to model selection, through the $\Delta\chi^2$ metric \citep[e.g.][]{perryman_astrometric_2014}, and even using the Bayesian information criterion as in \citet{ranalli_astrometry_2018} has its limitations. These limitations arise from the nature of the previous metrics, which solely rely on the maximum likelihood to be calculated. If the posterior distribution for the parameters is wide, skewed or even non-unimodal, this information fails to propagate into the preference criterion. Nested sampling gives access to the Bayesian evidences $P(D |\mathcal M)$ for each model $\mathcal M$
\begin{equation}
    P(D |\mathcal M) = \int_\theta P (D | \theta,\mathcal M) P(\theta|\mathcal M) d\theta
\end{equation}

The evidence represents the marginalisation over the parameter space $\theta$ of the likelihood $P (D | \theta,\mathcal M)$ and the prior $P(\theta|\mathcal M)$. This metric hence contains information about the entire parameter space, and not just the maximum likelihood. We can then calculate the Bayes factor. 
\begin{equation}
    BF = \frac{P(D |\mathcal M_1)}{P(D |\mathcal M_2)}
\end{equation}

The Bayes factor represents an odds ratio between the probability that model $\mathcal M_1$ is true given data $D$ and the probability that model $\mathcal M_2$ is true given the same data. It is hence a better metric to establish model preferences. As in \citet{kass_bayes_1995}, we consider that there is decisive evidence in favour of model $\mathcal M_1$ if $\log_{10}(BF)>2$ .

The second major advantage of nested sampling is that it can reliably find the absolute maximum of the posterior distribution, and map out non-Gaussian and multi-modal posteriors. This feature is especially useful to this study, as we aim to investigate model degeneracies. Specifically, if there are two sets of distinct solutions that produce similar likelihoods, say a 2-planet solution and a 1-planet solution, an MCMC routine can get stuck at the 1 planet solution (depending on the starting parameters) when the 2-planet solution can consistently be disambiguated with nested sampling.

We fit in terms of orbital parameters instead of Thiele-Innes constants, because of the nature of the fitting routine. A linear problem is useful in the case of MCMC fitting, but in our case it increases the size of parameter space to probe. Fitting in terms of orbital angles is non-linear, but this non-linearity is unimportant given the flexibility of nested sampling.

\section{Testing the pipeline for single planet detections} \label{sec:1_planet}

We demonstrate here the detection capabilities of our pipeline for single-planet systems, based on early data from Gaia DR4. We then recover the known single planet detection capabilities of Gaia through an injection-recovery grid study.

\subsection{Gaia BH3}

We aim to test the accuracy of our fitting pipeline with the only publicly available DR4 data to date, namely the Gaia BH3 data from \citet{panuzzo_discovery_2024}. 
Given that the object to detect far exceeds the mass prior space described in Table \ref{tab:priors}, we adjust the mass prior space upper limit to $q = 60$. The data itself contains the along-scan position measured on each of the 9 CCDs for every transit event, with the associated error. The mean error is $143\, \mu$as, which is significantly larger than the previously assumed $30\, \mu$as. However this smaller error is an estimate per transit. We hence combine the errors per transit by adding them in inverse quadrature ($1/\sigma_{\text{transit}}^2 = \sum_i^9 1/\sigma_{\text{CCD}_i}^2$), discarding the outliers (specified in the data by an outlier flag). Now the mean error per transit is $46 \, \mathrm{ \mu as}$, which is closer to the value assumed when generating mock data for this work. We still rely on additional processing of the full DR4 dataset (currently unpublished) to reach the ideal error value.

We compare our results with the results in \citet{panuzzo_discovery_2024} in Table \ref{tab:BH3}. We find that all retrieved orbital and astrometric quantities are consistent within the given uncertainties, and the uncertainties themselves are on the same order of magnitude, except for the mass. We attribute this difference to the fact that we do not account for the correlation between semi-major axis and mass ratio, and do not take RV data into account in our analysis.
     
\begin{table} 
\centering
\caption{Astrometric and Combined Solutions for Gaia BH3, from \citet{panuzzo_discovery_2024}. Additional data from nested sampling includes 1 sigma bounds.}
\label{tab:BH3}
\begin{tabular}{lcc}
\hline
\hline
Parameter & Combined & Nested Sampling\\
          & solution & solution\\
          \hline

$p$ [mas]                        & $1.6933 \pm 0.0164$ & $1.6808^{+0.0091}_{-0.0190}$\\
$\mu_{\alpha^*}$ [mas yr$^{-1}$] & $-28.317 \pm 0.067$ & $-28.328^{+0.073}_{-0.154}$\\
$\mu_\delta$ [mas yr$^{-1}$]     & $-155.221 \pm 0.111$ & $-155.194^{+0.121}_{-0.259}$ \\
$e$                              & $0.7291 \pm 0.0048$ & $0.7282_{-0.0114}^{+0.0052}$\\
$I$ [deg]                        & $110.580 \pm 0.095$ & $110.59^{+0.10}_{-0.22}$ \\
$\Omega$ [deg]                   & $136.236 \pm 0.128$ & $136.21^{+0.14}_{-0.30}$ \\
$\omega$ [deg]                   & $77.34 \pm 0.76$ & $77.85^{+0.17}_{-0.35}$ \\
$a$ [au]                       & $16.17 \pm 0.27$ & $16.61^{+0.33}_{-0.68}$ \\
$m_{\text{BH}} $ [M$_{\odot}$]   & $32.7 \pm 0.82$ &  $33.34^{+2.27}_{-2.54}$ \\

\hline
\end{tabular}
\end{table}

\subsection{Checking with Gaia-like data} \label{sec:dr4_notebook}

\begin{figure}
	\includegraphics[width=\columnwidth]{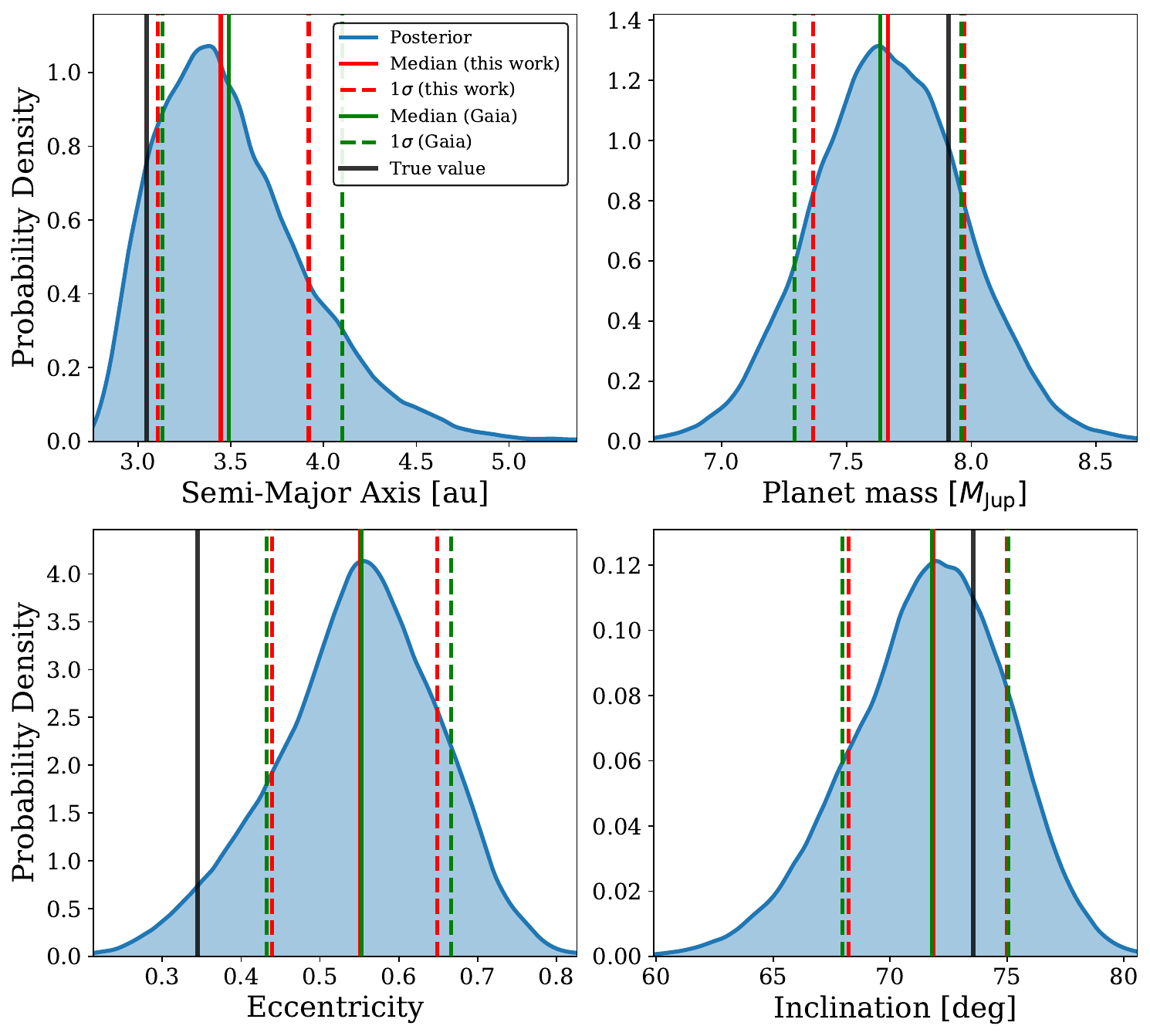}
    \caption{Posteriors in Semi-major axis, planet mass, eccentricity and inclination for the high SNR target (target 1). The true value, as well as the median and 1 $\sigma$ limits from nested sampling are also represented as vertical lines.}
    \label{fig:target_1}
\end{figure}

\begin{figure}
	\includegraphics[width=\columnwidth]{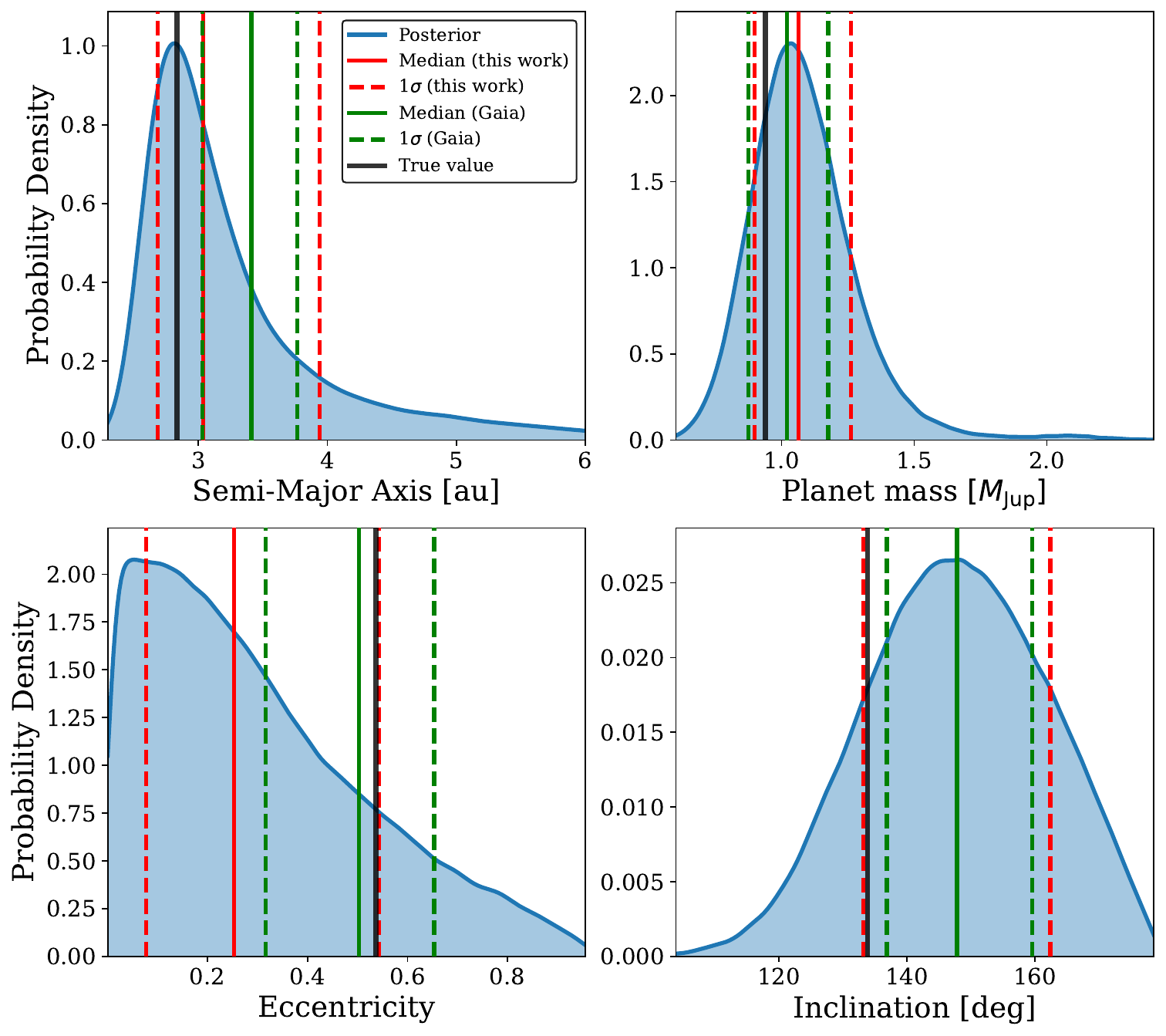}
    \caption{Posteriors in Semi-major axis, planet mass, eccentricity and inclination for the low SNR target (target 3). The true value, as well as the median and 1 $\sigma$ limits from nested sampling are also represented as vertical lines.}
    \label{fig:target_3}
\end{figure}

\begin{figure}
	\includegraphics[width=\columnwidth]{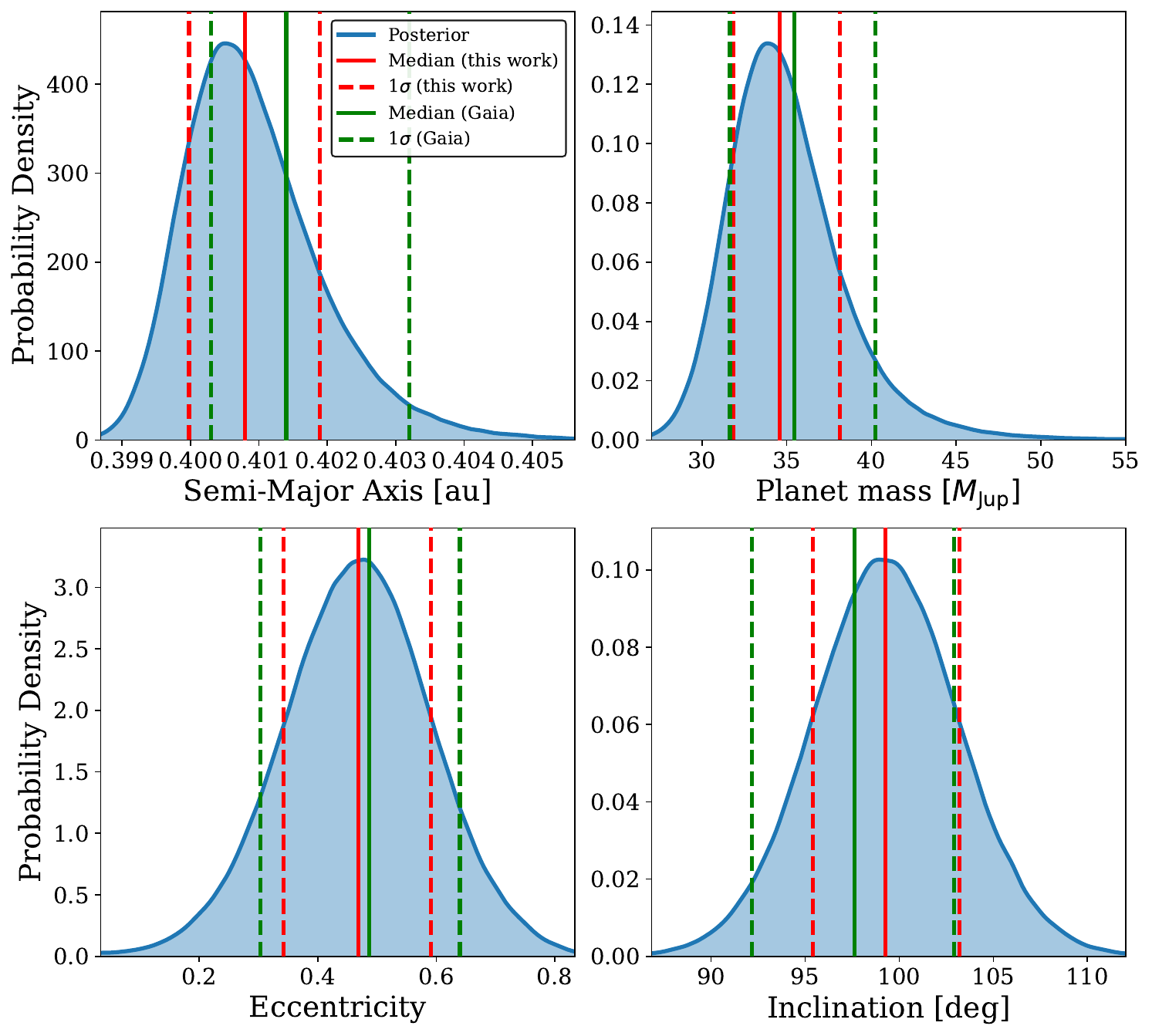}
    \caption{Posteriors in Semi-major axis, planet mass, eccentricity and inclination for the scan-angle related signal (target 4). The median and 1 $\sigma$ limits from the Gaia pipeline and from nested sampling are also represented as vertical lines in green and red respectively.}
    \label{fig:target_4}
\end{figure}

A source of mock data from Gaia DR4 can be found in \citet{delisle_dataset_2025}, which provides a set of real astrometric signals from DR4 in which single planet signals have been injected. Certain signals contain no planets, others contain long-period planets and low signal-to-noise signals. In this section, we aim to confirm that our pipeline can accurately run the model preference test, and compare the accuracy of our results with the Gaia planet retrieval pipeline approach (which uses a combination of periodograms and MCMC).

First we check that our pipeline's output results are consistent with the Gaia pipeline. To do this, we have compared the Bayes factor we get from our fit to their False Alarm Probability (FAP) in Table \ref{tab:BF_delisle}. The FAP is obtained from a time series analysis of the signal, and corresponds to the probability, under the null hypothesis (no planet), of obtaining a peak in the periodogram as significant as the one observed. We find good agreement between the two metrics, i.e. $BF < 1$ for non-detections ($\text{FAP} \sim 1$). We also note that both fitting routines misinterpret the scan-angle related signal as a planet; as a consequence, systematic errors are still an important factor to consider when fitting a planetary signal to the data. 

We show posteriors for semi-major axis, mass, eccentricity and inclination for targets 1, 3 and 4 in Figs. \ref{fig:target_1}, \ref{fig:target_3} and \ref{fig:target_4} respectively. Target 1 represents a large separation, high S/N planet, and in Fig. \ref{fig:target_1} we see a generally good agreement with the real values within $1 \, \sigma$. We hence infer for this ideal case with the 5 year data from DR4 that we can retrieve the semi-major axis to $\sim$0.5 au precision, the planet mass to $\sim 0.5 M_{\text{Jup}}$  precision and the inclination to a precision of $\sim 5$°. This tight precision in inclination can have important consequences in selecting a sample of Gaia planets for follow-up transits.
The solver seems to be less precise when it comes to measuring eccentricity, as the true value is outside of the $1 \, \sigma$ range, which itself has a width of 0.2. We explain the bias and wide spread in eccentricity by noise in the signal, the observation cadence for this specific target, and the general difficulty to get a tight constraint on eccentricity from astrometric data, especially for an irregular sampling frequency \citep[][]{holl_gaia_2023}. We compare this result to the Gaia pipeline analysis of the data. It yields $e = 0.55^{+0.11}_{-0.12}$, which is similar to our own outputs, hence confirming that the discrepancy is due to noise in the data. In order to find the true eccentricity it is hence necessary to take all aliasing effects into account. From these results we conclude that our pipeline is in good agreement with the Gaia pipeline, and accurately retrieves the orbital parameters. 

We draw similar conclusions for the case of the large separation, low S/N planet in Fig. \ref{fig:target_3}. We retrieve the mass to a $1 \, \sigma$ precision of $\sim 0.2 \, M_{\text{Jup}}$ and the semi-major axis to a precision of $\sim 0.5$ au. The lower resolution in the data does not seem to influence the precision at which these parameters are retrieved since they are directly related to the main frequencies of the orbit. However, the inclination is more weakly constrained (within $\sim 35$°). Eccentricity retrievals are the most affected by the low S/N of the data, as our model favours an eccentricity lower than 0.6, but the posterior is flat below $e$ = 0.5. The Gaia pipeline ($e = 0.52^{+0.19}_{-0.31}$) also provides a similarly weak constraint. 

We also show results for a scan-angle related signal (target 4), in Fig. \ref{fig:target_4}. We highlight this target to bring attention to \citet{holl_gaia_2023}, where low separation optical pairs and distant galaxy cores were found to induce a scan-angle dependent signal at certain specific periods. This signal contains no planets, yet both approaches confirm a planet detection to a high level of confidence. Since this is a period-dependent signal, the semi-major axis is extremely well constrained (to 0.002 au). The high precision on the retrieved semi-major axis, as well as the large planet mass (median at $\sim 35 M_{\text{Jup}}$) should hence be seen as tracers for an incorrectly identified signal. More work should be undertaken to provide approaches to filter out this false positive, specifically by including scan angle harmonics into the retrieval. In this study we encounter no further scan angle dependent signals, so this issue would only be relevant when applying the pipeline to real DR4 data.

\begin{table} 
\centering
\caption{Comparing the Bayes factors to FAP for mock DR4 observations }
\label{tab:BF_delisle}
\begin{tabular}{clcc}
\hline
\hline
Target & Comment & FAP & $\log_{10}(BF)$ \\
\hline

1                   & high SNR orbit $\sim$ 2000d   & $2.9\times10^{-38}$ &256\\
2        & single-star               & 1 & -0.8\\
3                & low SNR orbit $\sim$ 2000d   & 0.0018 & 7.9\\
4            & scan-angle related signal ($\sim$ 90d)               & $3.4\times 10^{-12}$ & 66\\
5                   & high SNR orbit $\sim$ 90d  & $4.9\times 10^{-21}$ &99\\
6 & low SNR orbit $\sim$ 90d & $0.00041$& 2.3 \\
7    &Long-period degenerate orbit  & $3\times 10^{-40}$& 334 \\

\hline
\end{tabular}
\end{table}

\subsection{Comparison to previous work and detection thresholds} \label{1pl_setup}

\begin{figure*}
	\includegraphics[width=\textwidth]{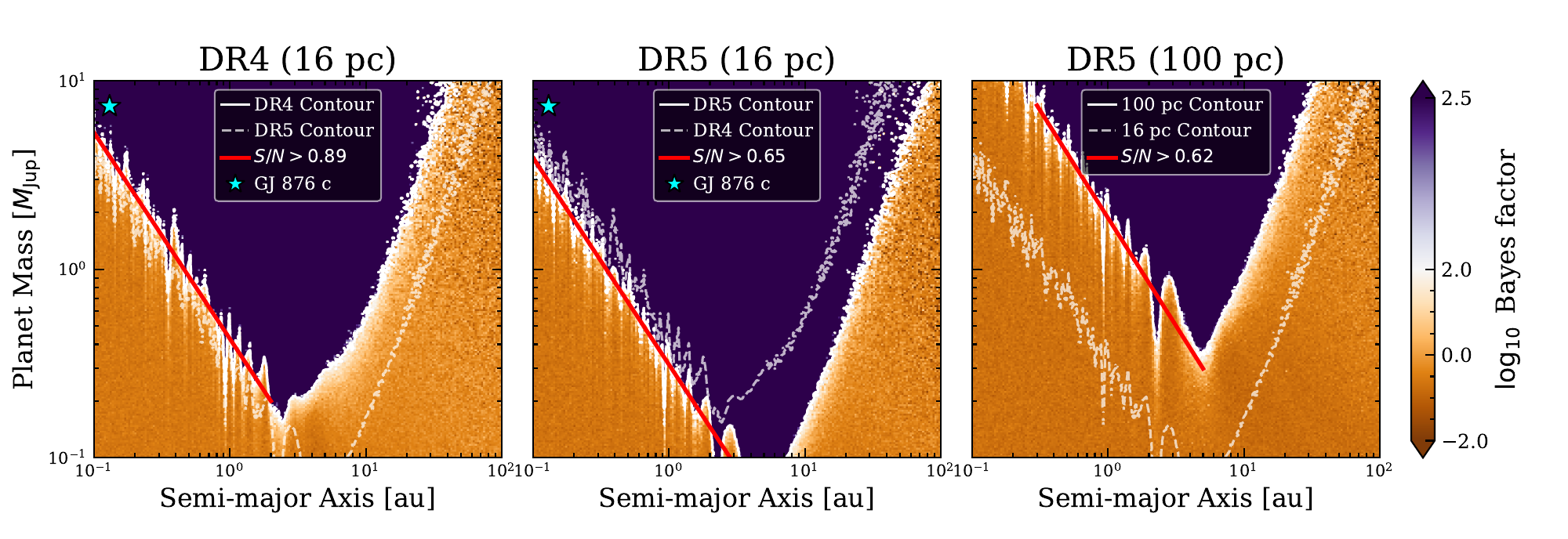}
    \caption{We find the Gaia detection space for planets using the Bayes Factor, for DR4 and DR5 data of systems at 16 pc and for DR5 data at 100 pc. We also show the empirical S/N detectability threshold for each case, assuming the systems are face-on. As an example, we show where planet c in the GJ 876 system (4.67 pc away) would lie, adjusting for distance by increasing the represented planet mass to offset the increase in distance to the system. Planet b is detectable, but would be represented beyond the range of the plot.}
    \label{fig:detection_space}
\end{figure*}

 We aim to provide a global characterisation of the detectability capabilities of our pipeline. To do this, we first run a parameter space study for the planet mass and semi-major axis (assuming a $1 M_\odot$ star). We generate a grid of 200 masses logarithmically spaced between 
 0.1 and 10$ \, M_{\text{Jup}}$ and 200 semi-major axes also logarithmically spaced between 0.1 and 100 au. These systems are face-on, all with $e=0.01$. We convert each orbit into an astrometric signal, using the recipe described in Section \ref{sec:generating1pl}. We then run our nested sampling pipeline to retrieve each signal, and present the resulting Bayes factor in Fig. \ref{fig:detection_space} for a single planet compared to no planetary signal. We consider that the planet is successfully detected if the Bayes factor provides decisive evidence in favour of the single-planet model.

There is a sharp turnaround in detectability at $\sim 3$ au and $\sim 5$ au ($\sim 5$ and $\sim 11$ years) for the DR4 and DR5 data, respectively, corresponding to the time covered by each of the final two Gaia mission data releases. This turnaround point separates the detection space in half by the semi-major axis. On the small semi-major axis side of the detection space we note that the boundary between the detectable and non-detectable regions is very clearly defined, and follows roughly a $m_{\text{p}} \propto a_\text{p}^{-1}$ trend for low semi-major axis on the left hand side of the plot. This trend is well understood \citep[e.g.][]{perryman_astrometric_2014,ranalli_astrometry_2018,lammers_exoplanet_2025} and is explained by the inverse relationship between the two quantities in Eq. \ref{eq:signature}. Matching the simulation data to the previous equation, we can deduce a S/N per astrometric point detectability threshold for DR4 and DR5 of 0.89 and 0.65 respectively, for the case of this specific region of space, and for face-on orbits. To establish a correspondence with the literature, we also recompute the grid study with randomised inclinations and $\cos I$ sampled uniformly between -1 and 1. When randomising inclinations, we find the expected increase of S/N detectability thresholds to 1.01 and 0.77 for DR4 and DR5 respectively. In the plots, we note the difference in detection capabilities between DR4 and DR5, as seen from the detectability contours. We also show results for an identical DR5 grid at 100 pc, and as predicted the parameter space is more restricted. We find very similar S/N per astrometric point thresholds between the 100 pc and the 16 pc simulations.

On the larger semi-major axis side of the detection space, the planet is harder to detect since Gaia is only able to collect data over a portion of the orbit, and the orbit is more poorly constrained. As the mass gets larger, the signal gets larger and our pipeline is able to confirm planets up to larger separations. This region should be treated with care because as we show in Section \ref{sec:acceleration} the inclusion of a linear acceleration term in the model without a planet might be preferred to a single planet fit. At larger semi-major axes, DR5 has a net advantage over DR4, allowing us to detect planets with periods up to twice as long for identical planet masses, and probe down to masses that are at least twice as low.

\begin{figure*}
	\includegraphics[width=\textwidth]{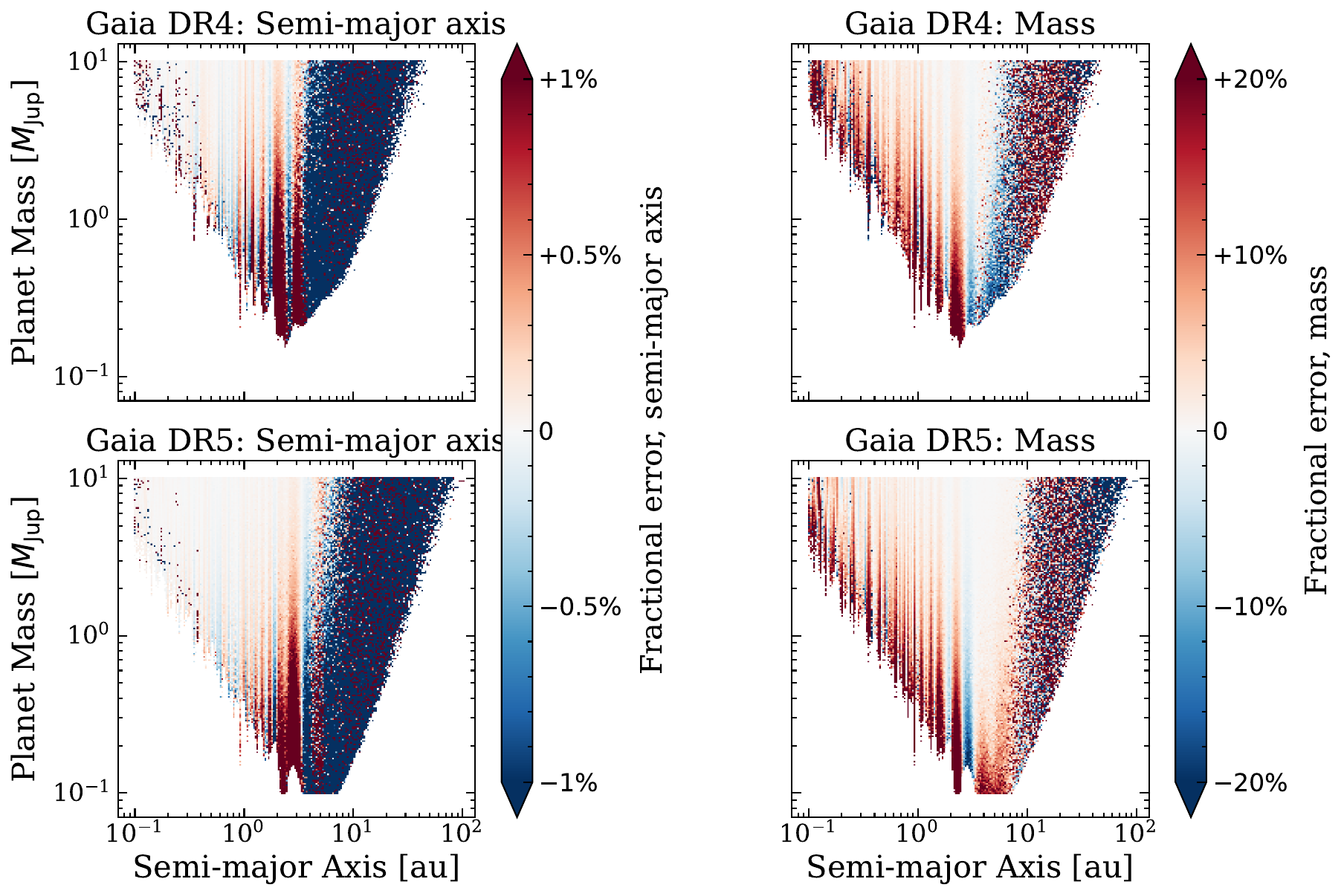}
    \caption{Fractional error on mass and semi-major axis for all systems at 16pc, as observed with DR4 and DR5, where only the systems with a confirmed detection are selected.}
    \label{fig:Condensed_Fractional_Error_Contours}
\end{figure*}

In Fig. \ref{fig:Condensed_Fractional_Error_Contours}, we show the fractional error on semi-major axis and mass for both DR4 and DR5 data. We see that 
for most planets within the detection region and with semi-major axes smaller than 7 au, we are able to constrain the planet's mass to better than 20\%. Beyond 7 au, we lose precision because we only detect a fraction of the orbit. We also notice vertical regions of constant semi-major axis where the error is larger, alternating between overestimation and underestimation of the mass. We relate this to the harmonics of other periodic signals, such as the Earth's revolution and the Gaia observation cadence, precession rate and rotation, biasing the retrieval at those specific frequencies. Within these strips, higher mass planets are more precisely measured. We bring attention to the shape of the contour and the slim vertical strip at 1 au, suggesting improved detection capabilities. In reality, these detections are false positives, as shown in \citet{holl_gaia_2023} and \citet{el-badry_generative_2024}. 

We also consider the error on the semi-major axis in Fig. \ref{fig:Condensed_Fractional_Error_Contours}. We note vertical regions of improved or hindered retrieved precision, directly corresponding to the vertical strips in mass highlighted previously. These errors, however, are not as strict as those in mass. We explain this difference by considering that Gaia can much better characterise periodic signals, and hence the semi-major axis. 

\begin{figure*}
	\includegraphics[width=\textwidth]{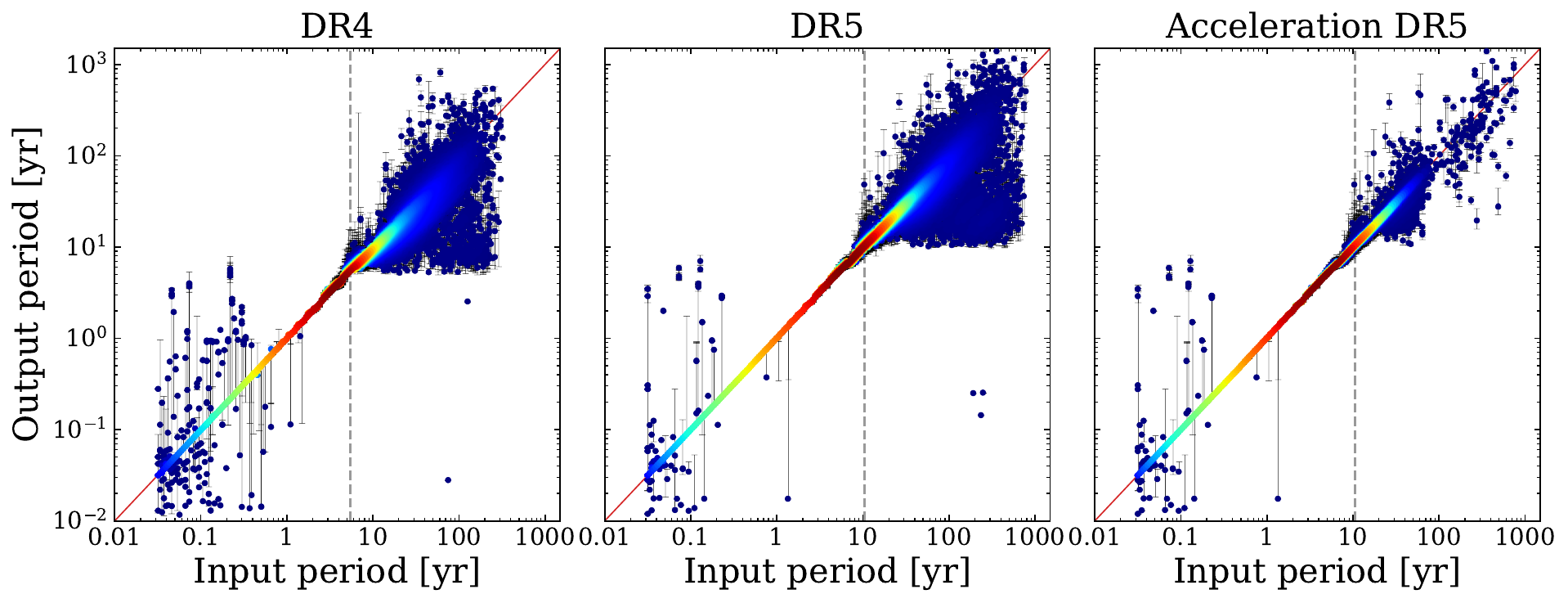}
    \caption{Comparing retrieved orbital period to input orbital period for the single planet case. The colour indicates the overlap density of points, with blue low and red high. Because of the grid structure of the parameter space, the density peaks at a few au, where the most planets are effectively detectable. In the rightmost panel we remove all systems that can be better explained with a linear acceleration in DR5.}
    \label{fig:periods}
\end{figure*}

We also assess the reliability of the retrieved period for DR4 and DR5 in the first two panels of Fig. \ref{fig:periods}. We find that for periods up to the data release time duration, the pipeline determines the planet's period to high accuracy, and tends to underestimate the period for orbits above this limit. For very short periods, we find spurious detections with misidentified periods, as a result of the choice of live points in the nested sampling. For individual real systems, especially at short periods where scan angle related biases can interfere, more care should be taken in the nested sampling resolution in order to better characterise the system. Beyond the respective data release baselines, more care must be taken in determining if the presence of a planet is statistically the best explanation for the signal. This is further explored in Section \ref{sec:acceleration}.

\begin{figure}
	\includegraphics[width=\columnwidth]{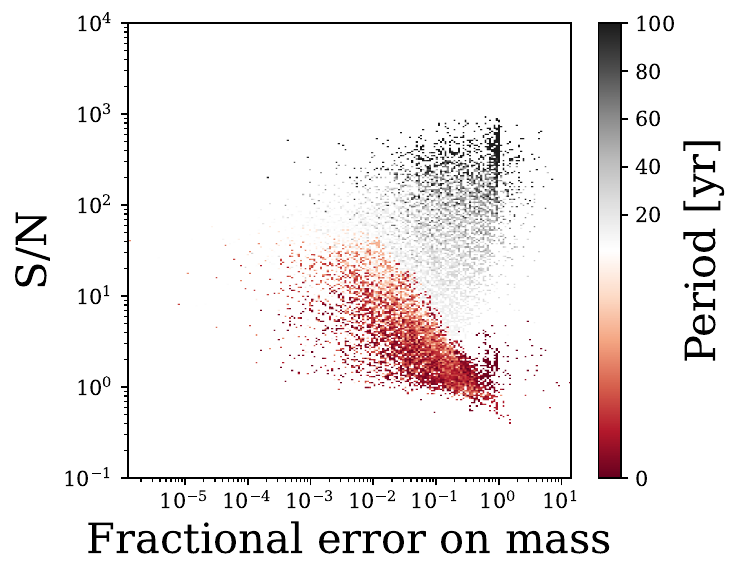}
    \caption{Planetary S/N for each detectable system of the parameter space study of DR4 at 16 pc, labelled by period. We represent the S/N with respect to the corresponding planetary relative error on mass. The colourbar shifts to grey beyond periods of 5 years.}
    \label{fig:1pl_accuracy}
\end{figure}

We furthermore find a clear trend in Fig. \ref{fig:1pl_accuracy}, where for all detected planets, lower fractional errors on mass are associated with the larger orbital periods (hence larger planetary S/N) as long as they are less than 5 years in DR4. After 5 years we see a turnaround. We find again that conversely, the longer period planets require a larger S/N per astrometric point to be detected, and if detected they present much larger errors on the mass. In fact, for the planets in the top right corner of the figure, given the long period, we need to investigate whether or not a planetary signal is statistically more significant than a simple linear acceleration.

\subsection{A linear acceleration: the case of long period planets} \label{sec:acceleration}

\begin{figure}
	\includegraphics[width=\columnwidth]{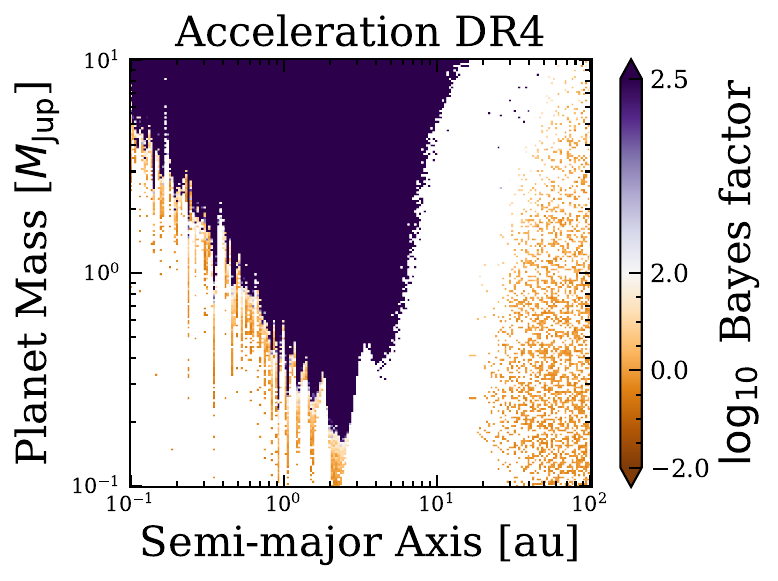}
    \caption{Detection space for planets in DR4 at 16pc. Systems for which the acceleration model is preferred are removed and shown in white}
    \label{fig:accel}
\end{figure}

When considering the results in Fig. \ref{fig:detection_space}, we are drawn to the long period planets detected at high significance, especially beyond 10 au. Given the Gaia mission lifetime of roughly 10 years, these planets would be observed for a fraction of their orbit only. For instance, a planet at 30 au around a solar-mass star would only be observed for 6\% of its orbit. To confirm that the signal is indeed from a planetary companion, we need to detect an elliptical periodic reflex motion. We therefore aim to assess the significance of such a detection with our pipeline. We add an extra 7-parameter astrometry model:

\begin{equation} \label{astrometry_accel_eq}
    \begin{aligned}
        &\alpha^* (t) = \alpha_0^* +\Delta\alpha_0^* + \mu_{\alpha^*} (t-t_0) + \frac{1}{2}\dot{\mu}_{\alpha^*} (t-t_0)^2 + \Pi_{\alpha^*} p\\
        &\delta (t)= \delta_0 +\Delta\delta_0 + \mu_\delta (t-t_0) + \frac{1}{2}\dot{\mu}_\delta (t-t_0)^2 +  \Pi_\delta p
    \end{aligned}
\end{equation}

which consists of Eq. \ref{astrometryeq}, to which we have added $\dot{\mu}_{\alpha^*}$ and $\dot{\mu}_\delta$, which are linear accelerations in right ascension and declination respectively. We generate a 200 by 200 grid of single planet signals, identical to the one in Section \ref{1pl_setup}. We then fit the data with 3 separate models: the 5 parameter astrometry model, the 7 parameter astrometry model and the 5+7 parameter single planet signal. In the case of the 7 parameter model, we set a uniform prior between $-10^{-4}$ and $10^{-4} \,  \text{as}\, \text{yr}^{-2}$, which corresponds to the range of accelerations of the longer period planets. 

We present the results for DR4 in Fig. \ref{fig:accel}. We find that beyond 6 au for all masses, the acceleration model is preferred for a large majority of systems. For DR5, a similar steep detectability barrier emerges around 10 au. In other words, the curvature of the orbit is not significant enough to infer the presence of a planetary companion. As a consequence, fitting a planetary signal will result in large uncertainties in orbital parameters, that will ultimately be prior dominated. This coincides in Fig. \ref{fig:Condensed_Fractional_Error_Contours} with the larger au region where the fractional error sharply transitions to larger values. Furthermore, in Fig. \ref{fig:periods}, when removing systems where acceleration is the preferred solution, we find that we also remove most systems that present a systematically lower retrieved period. These results motivate us to only consider a region of parameter space up to 10 au when studying multiple systems in the following sections.

\section{Multiple-planet detections} \label{sec:multiple_planet_detections}

Having built our planet detection and Bayesian model selection pipeline for a single planet, we now extend it to multiple planets. While the framework is fully generalisable to many planets, here we just focus on modelling the detection of two planet systems. 

\subsection{The setup} \label{2pl_setup}

We generate the astrometric signal from two planets using the same methods as in Section \ref{sec:generating1pl}, and specify the masses and initial semi-major axis of both planets.

The detection pipeline for the case of two planets is similar to the one described for a single planet in Section \ref{sec:model1pl}. We assume non-interacting Keplerian orbits for both companions, so that the final signal is a superposition of the signal described in Eq. \ref{eq:XY} generated by each planet:

\begin{equation} \label{eq:XY_2pl}
    \begin{aligned}
     &X_{\text{2pl}}(t)= A'_1x_1(t)+ F'_1y_1(t) + A'_2x_2(t)+ F'_2y_2(t)\\
     &Y_{\text{2pl}}(t)=B'_1x_1(t)+ G'_1y_1(t) + B'_2x_2(t)+ G'_2y_2(t)
    \end{aligned}
\end{equation}
where the subscripts indicate the orbital elements  for each planet. The resulting signal then requires a 19-parameter fit.

We use nested sampling to fit the data, with the same likelihood model as in Eq. \ref{eq:likelihood}. The priors for each orbital parameter are identical to those described in the case of a single planet, as shown in Table \ref{tab:priors}. Here, we specifically use nested sampling to circumvent any fitting degeneracies, and to extract a Bayes factor, allowing us to run a model preference test between the cases of 0 planets, 1 planet and 2 planets. We found that using 2500 live points is sufficient on the scale of our overall parameter study.

\subsection{Can Gaia detect 2 planet systems? The example of GJ 876} \label{sec:2planet_example_876}

\begin{figure*}
	\includegraphics[width=\textwidth]{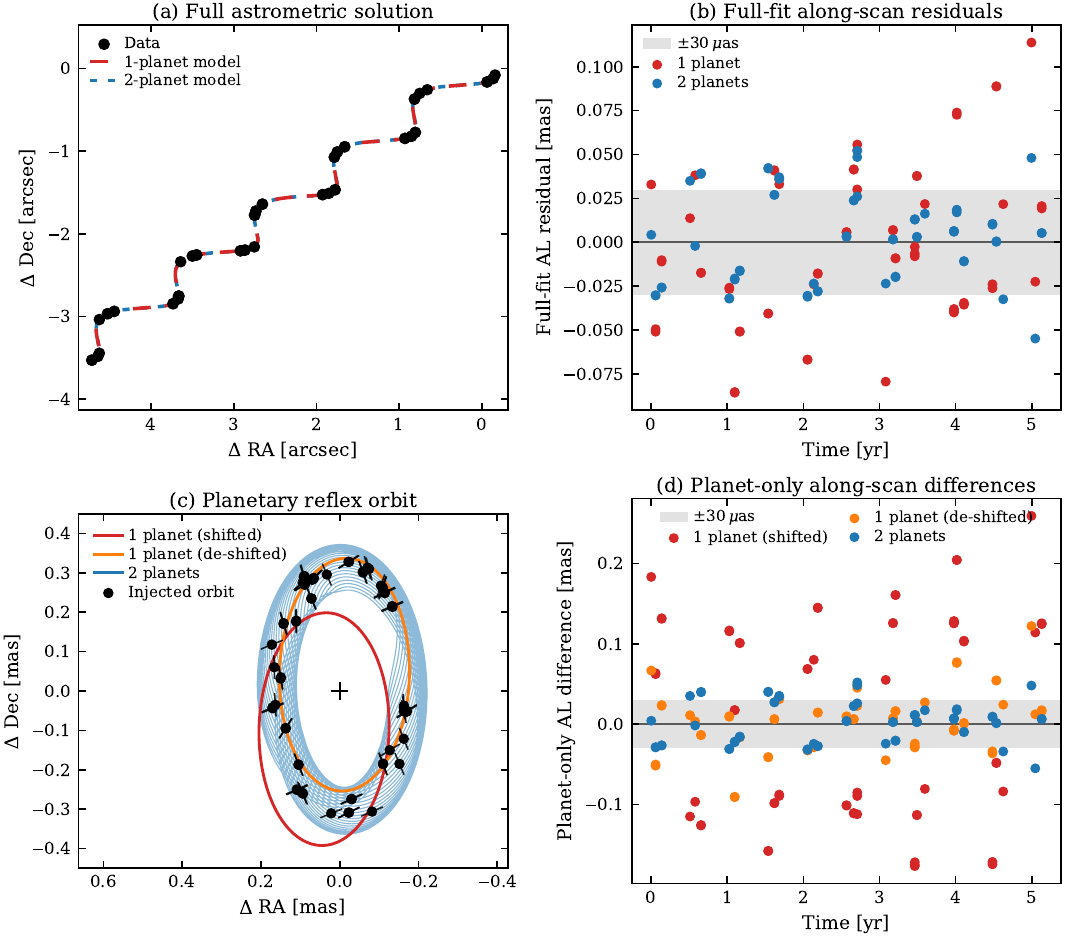}
    \caption{ Fits to our simulated GJ 876 data with planets in dynamical resonance. Top left: reconstructed astrometric path in the sky, with both 1-planet and 2-planet models superimposed. Top right: Along-scan residuals for the full astrometric fits, for both models. The assumed 30 $ \mu$as astrometric resolution is also represented as a grey band. Bottom left: Comparing the two models with the astrometric data, with parallax and proper motion subtracted. The 1-planet fit gets shifted, hence we also represent the de-shifted orbit, better fit to the subtracted data. Bottom right: Residuals for the reflex motion only. The shifted and de-shifted residuals are represented for the 1-planet fit.}
    \label{fig:dr4_gj876_fits}
\end{figure*}

It has been shown that Gaia has the capabilities to detect multiple planets \citep[][]{sozzetti_detection_2001,casertano_double-blind_2008}.
We will also demonstrate that Gaia is able to detect multiple planets, by first investigating a real system known to host planets in 2:1 resonance. We then extend this idea to a parameter space grid of two-planet systems.

To date there are no publicly available astrometric data for a multi-planet system. We hence choose to simulate this data. We choose to simulate the well-studied system GJ 876, because it has known and precisely determined orbital parameters \citep[][]{rivera_lick-carnegie_2010,nelson_empirically_2016,rosenthal_california_2021}, planets b and c are confirmed to be in dynamical resonance \citep{marcy_pair_2001, laughlin_short-term_2001}, and expected to be theoretically detectable in Gaia astrometry, as demonstrated in Fig. \ref{fig:detection_space}. Additionally, this system has published fits from DR3 data \citep{holl_gaia_2023}.

To simulate the system, we use ReboundX \citep{tamayo_reboundx_2020} and generate a dynamical copy of GJ 876 b and GJ 876 c by setting up the planets out of resonance, and applying the migration prescription and eccentricity damping parameters from \citet{lee_dynamics_2002}. This leads to the planets librating in dynamical resonance with semi-major axes and eccentricities consistent with observations. We also integrate the system until it is in a state that matches the known measured ephemeris from \citet{rivera_lick-carnegie_2010}. We then use the known proper motions and parallax of the system to generate a realistic astrometric observation. We also use GOST to find the observation times and scan angles for when Gaia has observed the system. We keep the assumed 30 $\mu$ as error per astrometric data point, given that the star has a G-band magnitude of 8.88 \citep{bruno_detailed_2024} which falls into the most optimal magnitude region for Gaia observations \citep[][]{lindegren_gaia_2021, lammers_exoplanet_2025}. We then fit the resulting signal with 1-planet and 2-planet models, for the observation baseline of DR4. We show our fit to the synthetic data in Fig. \ref{fig:dr4_gj876_fits}.

From the fit, we find that the 2 planet model is preferred, with a Bayes factor of $\log_{10} \text{BF}_{21} = 6.3$, where the subscript indicates that we take the ratio of the evidences between the 2-planet and the 1-planet models. Visually, on the reconstructed fit of the motion on the sky we find both models to be indistinguishable; however, the AL fit residuals can disambiguate model fits. We find an RMS of 44 $\mu$ as and 25 $\mu$ as with a reduced $\chi^2$ of 2.7 and 1.0 for the 1-planet and 2-planet model respectively, hence confirming that the 2 planet model is preferred.

In the bottom two panels of Fig. \ref{fig:dr4_gj876_fits}, we show the planetary orbit fit only. The 1-planet solution as found from the solver (in red) is naturally shifted, and shows very large residuals (125 $\mu$as). This shift will be of importance later when discussing orbital degeneracies. The shift arises from the resonant configuration of the planets. Since the 1-planet fit returns a moderate eccentricity of 0.36, the barycentre must be shifted for the fit to align with the reflex motion data once the proper motion and parallax are subtracted. For now, we assume that the shift is carried in the $\Delta \alpha_0^*$ and $\Delta \delta_0$ parameters. We hence artificially de-shift the 1 planet orbit by shifting the celestial coordinates to minimise the RMS for the reflex motion only. The de-shifted fit is a much better fit to the signal than the native shifted fit. However, comparing the de-shifted 1 planet and 2 planet AL residuals (37 $\mu$ as and 25 $\mu$ as respectively) we find again that the 2-planet model is statistically preferred.

\subsection{Two planet detection parameter space} \label{sec:grid_study}

Having demonstrated that Gaia is capable of detecting real multiple exoplanet systems, we want to probe the detection capabilities of \textit{Gaia} for two planet systems with a parameter study. We set up a $50 \times 50$ grid of planets in semi-major axis and mass, spaced logarithmically between $0.1$ and $10\,\mathrm{au}$, and $0.1$ and $10\,\mathrm{M_\text{Jup}}$, respectively. For each planet on the grid, we generate a second planet. Given the scope of this study, we focus on planets near the $2{:}1$ resonance. Hence, for each planet in the grid, we associate a second planet with period ratios $P_\text{rat} =P_\text{out}/P_\text{in} = P_2/P_1 $ of $1.7{:}1$, $1.9{:}1$, $1.95{:}1$, $2{:}1$, $2.05{:}1$, $2.1{:}1$, and $2.3{:}1$. This is to ensure that we get an accurate characterisation of our inference framework as we approach commensurability. We also consider secondary planets with mass ratios $m_\text{rat} = m_\text{out}/m_\text{in} = m_2/m_1$ of $0.2$, $0.4$, $0.8$, $1$, $1.5$, $2.5$, and $5$, neglecting any systems where the planet is above $10\,\mathrm{M_\text{Jup}}$ or below $0.1\,\mathrm{M_\text{Jup}}$. We execute these cut-offs to avoid any unnecessary computations considering systems clearly below the one-planet detection limit, and to constrain the study to giant planets only. This results in $102\,600$ systems.

\subsection{Two planet S/N detection criteria}

\begin{figure*}
	\includegraphics[width=\textwidth]{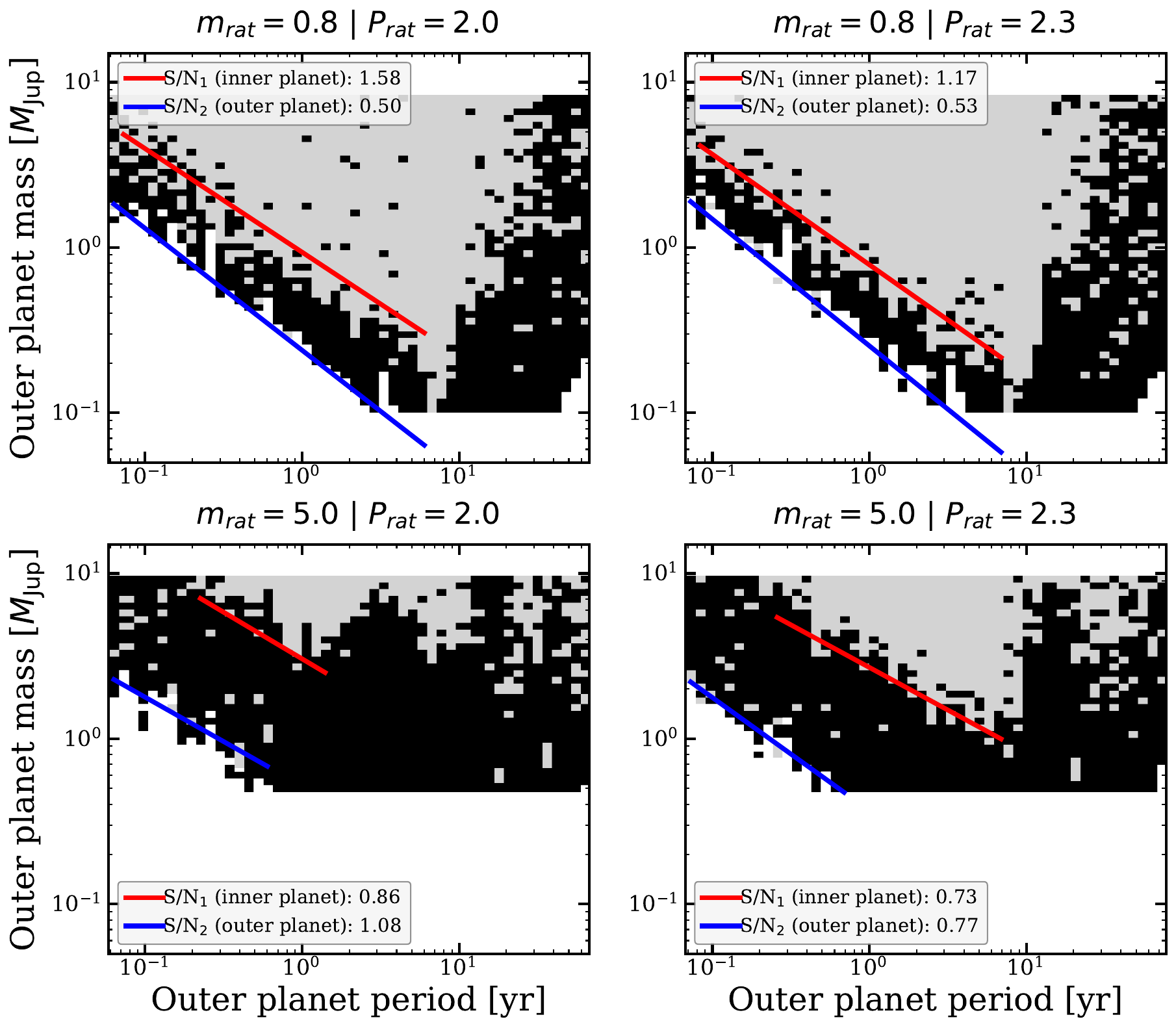}
    \caption{1 and 2-planet detectability thresholds for 4 specific configurations of planet mass and period ratios (close to and outside of resonance). The black grid points represent systems where only a single planet is detectable. Grey points show where both planets are detectable. The red line is a power law fit to the upper black contour, parametrising the 2-planet detection threshold. Similarly, the blue line is a power law fit to the lower contour and parametrises the 1-planet detection threshold.}
    \label{fig:2pl_SN_example}
\end{figure*}

\begin{figure*}
	\includegraphics[width=\textwidth]{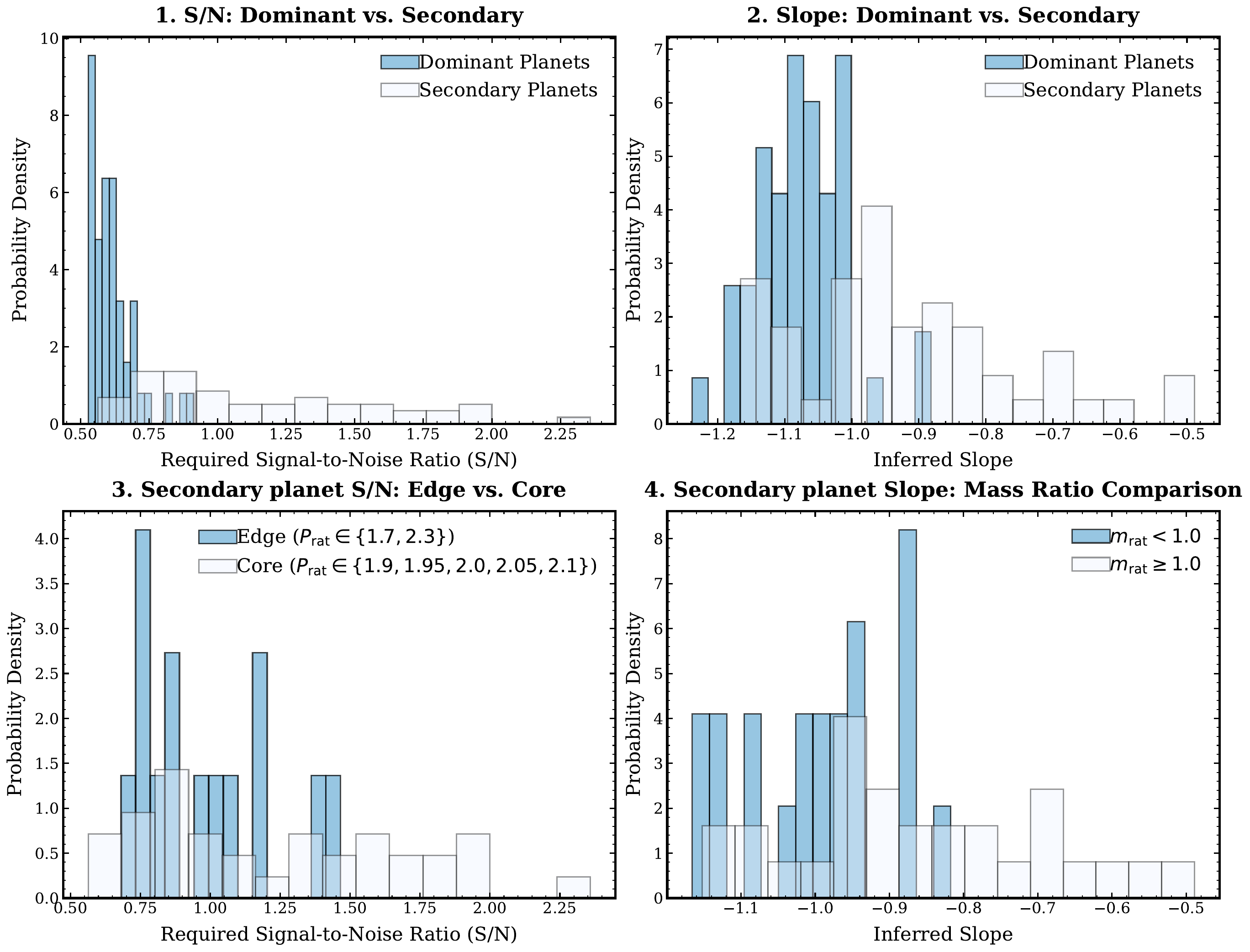}
    \caption{Multi-planet system detection diagnostics, from all 49 planet mass and period ratio configurations of the grid study in Section \ref{sec:grid_study}. Top left: fitted S/N detection threshold, separated between dominant planet in the signal and secondary planet. Top right: slope of the fitted power-law detection boundary, also separated out between dominant and secondary planet. Bottom left: Amongst secondary planets, S/N detection threshold, separated between systems away from (1.7:1 and 2.3:1) and near resonance (all other period ratios). Bottom right: slope of the fitted power-law detection boundary, also separated between systems with low ($m_{\text{rat}} < 1$) and high ($m_{\text{rat}} \geq 1$) planet mass ratios.}
    \label{fig:paper_combined_plots}
\end{figure*}

Similar to the single planet case, we want to characterise the S/N per astrometric point detection criteria for the 2 planet case, when fitting with a 2-planet model. In Fig. \ref{fig:2pl_SN_example}, we isolate in black the systems where only a single one of the two planets is statistically detectable. We colour the systems where both planets are detectable in grey. There are 4 plots, sampling the larger parameter space, in different planet mass ratios and period ratios. On the left-hand side of each plot, we can fit a power law to the upper and lower edges (represented in red and blue)
\begin{equation}
    m_\text{p} = k a_\text{p}^s
\end{equation}
where $k$ is a constant that can be related to the S/N detection threshold using Eq. \ref{eq:signature}, and $s$ is the slope of the line in log-space (it should be -1 if the S/N detection criterion follows from Eq. \ref{eq:signature}). We isolate the systems where only 1 planet is detectable in black since the lower limit of this detection space traces the detection limit for the planet with the highest S/N (in the cases presented here the outer planet, in blue). The upper limit traces the S/N detectability of the planet that is harder to detect (here the innermost planet, in red). The outer (most detectable) planet here acts as a proxy to connect with our conclusions for single planet systems. We want to know whether or not the single planet detection criterion also applies to each individual planet in a 2 planet signal, or if the joint signal requires adjustments to the criterion. We find that the S/N detectability threshold for these planets is still in relatively good agreement with the theory S/N $\sim$ 1. The less detectable inner planet often requires a higher S/N threshold to detect, due to the additional parameters in the fitting model.


The case of planets in 2:1 commensurability with high planet masses is particularly interesting (lower left panel). We see that there is a bump between 1 and 4 years where the 2 planet system is not detectable, stopping the red line from providing a meaningful constraint beyond 1 au. 
In reality, when considering planets in exact dynamical resonance as in Section \ref{sec:librations}, we find that the edge fit by the red line completely disappears in favour of all systems being detected as a single planet system. We explain this discrepancy by considering that at short periods, the planets experience more conjunctions over the observational baseline. If they are not in dynamical resonance, the longitude of the conjunction will drift and the signal from each individual planet can be distinguished. In resonance, the longitude of the conjunction librates and hence for each orbit the orbital shape repeats itself with the same features at the same orbital longitudes, hence hiding the observational signature of the inner planet. This is a clear case where the single-planet detection criterion fails in a significant manner.

To draw a more robust systematic conclusion for the 2-planet S/N detection criteria, we present 4 histograms in Fig. \ref{fig:paper_combined_plots}. We present distributions of S/N detection threshold and slope $s$ over all 49 configurations of mass and period ratios for the 2 planets. 

We start with the S/N detection threshold. In the upper left panel of Fig. \ref{fig:paper_combined_plots}, we find that the dominant planet in the signal almost consistently has a S/N detection threshold between 0.5 and 0.75.  For the less dominant planet, we find that 0.5 < S/N < 2. This preference is due to the added parametric complexity of a 2-planet model.
We can further ask whether there is a criterion that leads to higher or lower detectability thresholds in the lower signal planet.

There is no significant dependence of the secondary planet S/N threshold on the mass ratio. However when the planets are closer to 2:1 commensurability, as represented in white in the lower left panel, this S/N threshold is larger. On the contrary, when the planets are away from resonance, as represented in blue in the same panel, the S/N threshold values are lower (<1.5), but still above the threshold value found for the dominant planet. We run an Anderson-Darling test on the two distributions, and find them to differ with a confidence of 90 \%, showing weak evidence that resonances play a role in hindering multi-planet detections. Further analysis in the following sections will show stronger numerical and analytical evidence for this hindering effect.

The second metric, the inferred slope of the power law, is a proxy for how accurately the model in Eq. \ref{eq:signature} represents the S/N thresholds. From the top right plot, we see that this slope for the dominant planet is bounded by $-1.2 < s < -0.9$, close to its theoretical value of -1. For the case of the secondary planet, the bound upper value can deviate up to -0.5.
Finally, in the bottom right plot of Fig. \ref{fig:paper_combined_plots}, we find that, for the less detectable planet, most deviations in the slope are carried by systems with larger outer planet masses. 

Overall, and unsurprisingly, we find that the lowest S/N planet will be consistently harder to detect. The largest deviations in the detectability criteria are found when the planets are close to 2:1 commensurability, and when the outer planet is more massive than the inner planet.

\subsection{Biases due to multiple planets and resonances}

\begin{figure*}
	\includegraphics[width=\textwidth]{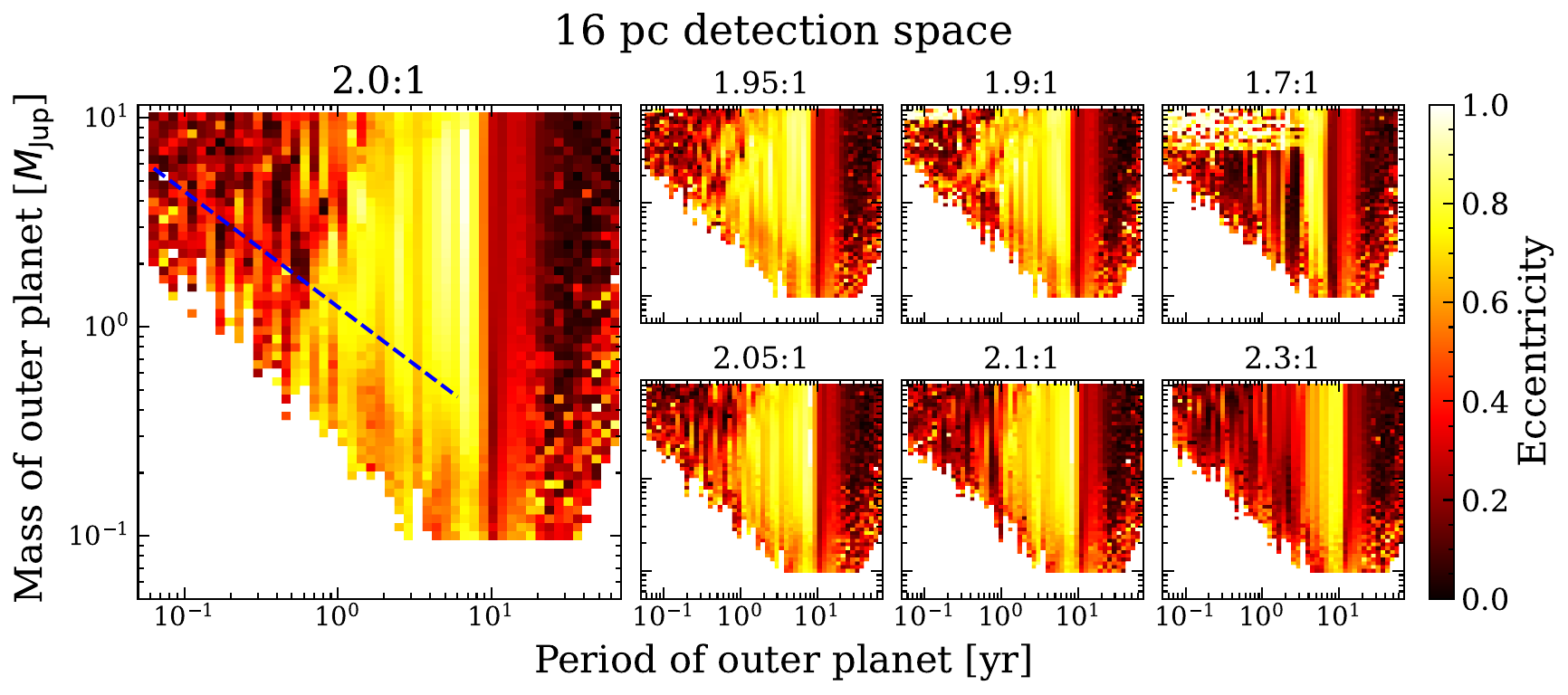}
    \caption{Effective eccentricity obtained from fitting a 1-planet model to data from a circular 2-planet system where the planets are equal in mass, as a function of outer planet mass and orbital period, for the DR5 baseline. All period ratios from the grid study are represented. Each non-white grid pixel represents one system, where at least 1 planet is detectable. The blue line is the power law fit to the 2-planet detectability contour, as defined by the Bayes factor.}
    \label{fig:money_plot}
\end{figure*}

\begin{figure}
	\includegraphics[width=\columnwidth]{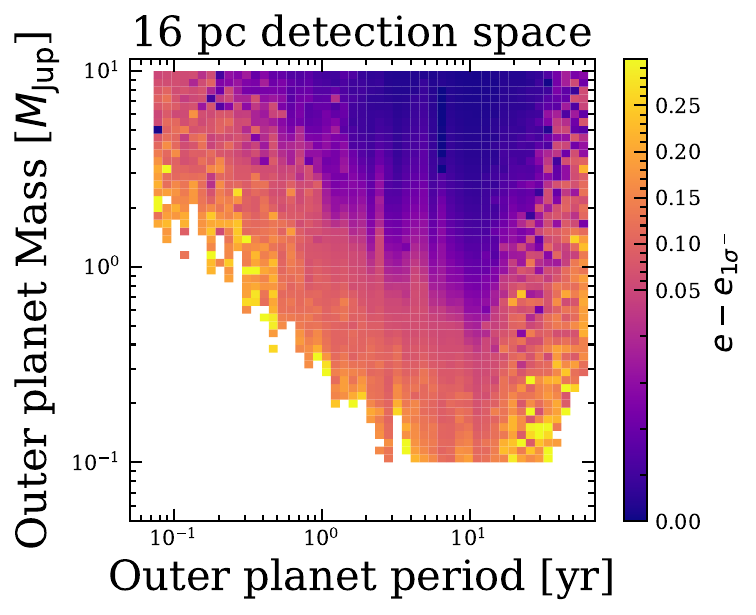}
    \caption{1-$\sigma$ lower error on the effective 1-planet eccentricity represented in the leftmost panel of Fig. \ref{fig:money_plot}, where both planets in the signal are initialised on circular orbits. We find that the uncertainties are approximately symmetric; therefore, this figure is similar to one obtained by considering the 1-$\sigma$ upper error.}
    \label{fig:money_plot_error}
\end{figure}

\begin{figure*}
	\includegraphics[width=\textwidth]{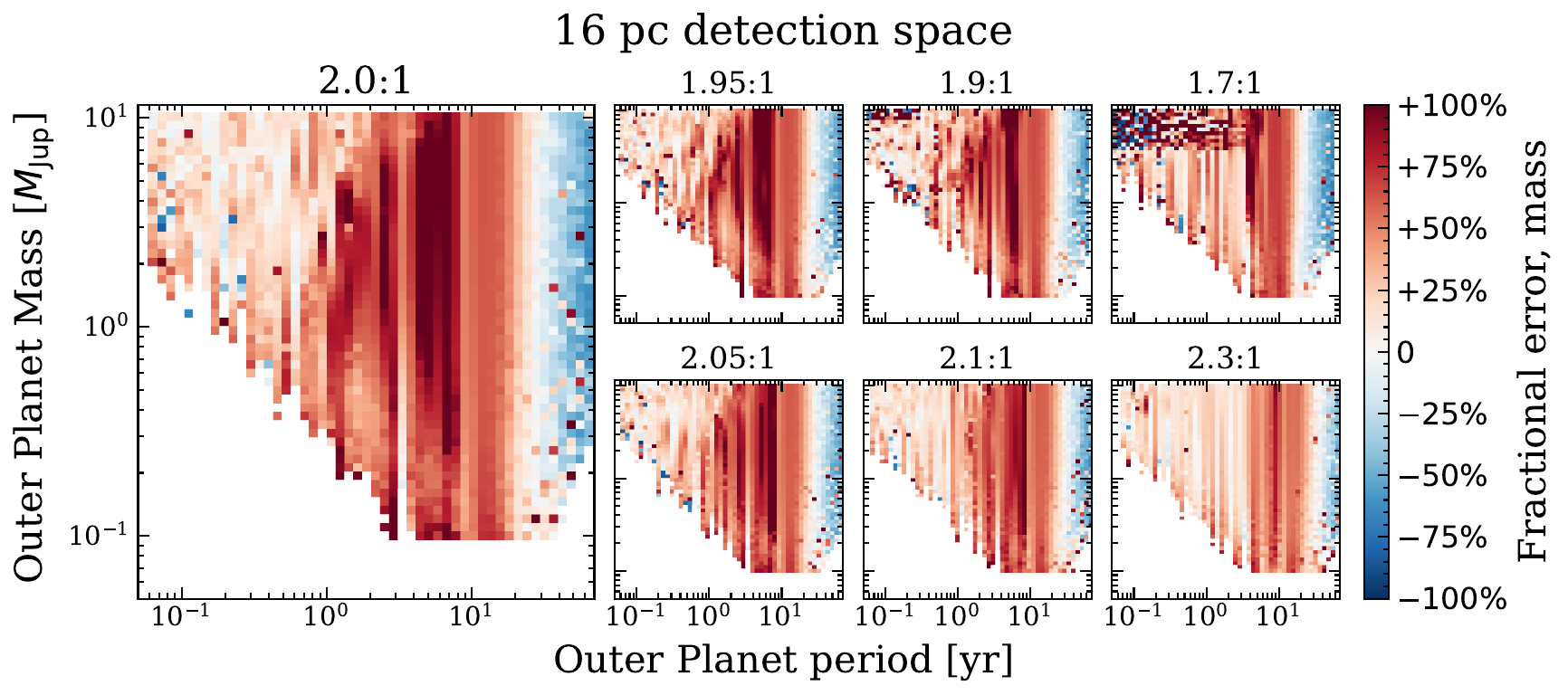}
    \caption{Mass fractional error obtained from fitting a 1-planet model to data from a circular 2-planet system where the planets are equal in mass, for the DR5 baseline. All period ratios are represented. Each grid pixel represents one system. The blue line is the power law fit to the 2-planet detectability contour, as defined by the Bayes factor.}
    \label{fig:money_plot_mass}
\end{figure*}

When fitting a single planet model to a two planet signal, we would ideally expect to recover the planet with the strongest S/N. In practice, the extra signal due to the second, less detectable planet is at risk of contaminating the fit for a single planet. We hence aim to characterise the extent of this contamination on the significance of any single planet detection, and more importantly on the inferred orbital parameters for this planet. 

We focus on the retrieved orbital parameters obtained by fitting a single-planet model to the two-planet signals in our grid study. Both planets in the data have near zero eccentricity, hence this should also be reflected in the single planet fit. In Fig. \ref{fig:money_plot}, however, we find that this is not the case. We select the simulations where the planets are equal in mass, and show the retrieved eccentricity for the 1 planet fit, across the different period ratios for DR5. When the planets are near commensurability for a 2:1 resonance, there exists a range of outer planet periods between 0.6 and 10 years where the inferred eccentricity is consistently found to be greater than 0.6. When the planets are further away from resonance, this feature does indeed seem to become less relevant across the parameter space as a whole. When we consider planets with 1.9:1 or 2.1:1 period ratios, the high eccentricity region lower limit slowly moves to 1 year and for 1.7:1 or 2.3:1 period ratios, this high eccentricity region still persists but is only relevant between periods of 4 and 7 years. 

We confirm that this feature is indeed a systematic error by looking at the error on the eccentricity, as shown in Fig. \ref{fig:money_plot_error}. We find a strong overlap between the region of high eccentricity previously found and the region of low error, below 0.1. We hence conclude that the posteriors are narrow around these large, unphysical eccentricities, and that the solver finds a valid high eccentricity orbit, hence the planet with the lower S/N does indeed contaminate the single-planet fit in a significant way.

Furthermore, we identify four regions of distinct eccentricities in Fig. \ref{fig:money_plot}, for planets in 2:1 commensurability. For short period, high mass planets, Gaia is able to measure enough planetary orbits to cover the shift in the alignment of the signal for each orbit. The large masses lead to planet-planet interactions that also break this alignment. This results in low effective eccentricities. There is a rapid shift as the outer planet period increases to $\sim 1$ yr.

The second region, as previously discussed, is the region with the highest effective eccentricity for outer planet periods between 0.6 and 10 yrs. However, we can make out a third region, below the blue line in Fig. \ref{fig:money_plot}. The line is a power law fit to the 2 planet detectability contour. Below the line are systems where the planet masses are lower, hence the inner planet does not have the required S/N to be detectable. We also find that despite another planet not being statistically detectable, the single retrieved planet is found to have a significant eccentricity, albeit lower than the one for systems above the blue line. This biased eccentricity will have consequences on the level of cold Jupiter demographics from Gaia, where large eccentricity giant planets are likely to be overrepresented. Other detection methods such as RV follow-ups \citep[e.g.][]{barbato_gaps_2026,pinamonti_gaps_2026,marcussen_radial_2026} would be required in order to resolve this degeneracy. The distinction between the two regions of 1 or 2 planets being detectable is of particular interest to us and will be investigated in subsequent sections.

The final region is for any system with the outer planet period beyond 10 years, i.e. the Gaia DR5 baseline. Gaia measures less than one outer planet orbit, hence the orbits do not repeat and there is no eccentricity bias.

In Fig. \ref{fig:money_plot_mass}, we report another bias in the retrieved planet parameters. The plot shows the fractional error between the retrieved mass for the single planet fit and the theoretical mass of the outer planet. Since the outer planet has the strongest signal, we expect to retrieve it with the 1-planet fit. In the exact same parameter space where we found the eccentricity bias, we find that the outer planet mass fractional error is heavily biased and consistently overestimated by over 50 \%. The four regions of distinct eccentricity discussed previously can be mapped onto this plot as regions of mass bias. The bias can lead to retrieved planets with twice the mass of the planet responsible for the signal.
We similarly find that this bias weakens as the planets are further away from commensurability. Additionally, beyond 30 years, when the outer planet signal becomes statistically similar to that of a linear acceleration, the planet mass is underestimated. Since the inner planet has the same mass as the outer planet, this means that although the inner planet might be detectable and distinguishable from a simple linear acceleration, the added (unconstrainable) signal from the outer planet biases the retrieval. In both Figs. \ref{fig:money_plot} and \ref{fig:money_plot_mass}, we want to highlight that the high mass systems in the 1.7:1 period ratio are susceptible to dynamical instabilities, and generate noise in the upper left of the plot. To accurately study these systems, we would need to integrate these systems for Gyrs to remove the unstable systems; therefore, we do not trust our results in this region of parameter space.

\subsection{From commensurability to dynamical resonance} \label{sec:librations}

So far, we have made the assumption that the signal from planets near the 2:1 commensurability makes no difference in the retrievals if compared to that of planets in dynamical mean motion resonance.

In Figs. \ref{fig:money_plot} and \ref{fig:money_plot_mass}, in the 2:1 case we find in the upper left corner (short period and high mass) that the eccentricity and mass biases are absent or very weak. In this regime, many planetary orbits will be observed by Gaia, and the more massive planets produce stronger planet-planet interactions. These interactions cause deviations from the exact alignment of each orbit and the degeneracy breaks (i.e. non-interacting Keplerian orbits are poor approximations). We decide to investigate this further by setting up the planets in actual dynamical resonance. 

\begin{figure}
	\includegraphics[width=\columnwidth]{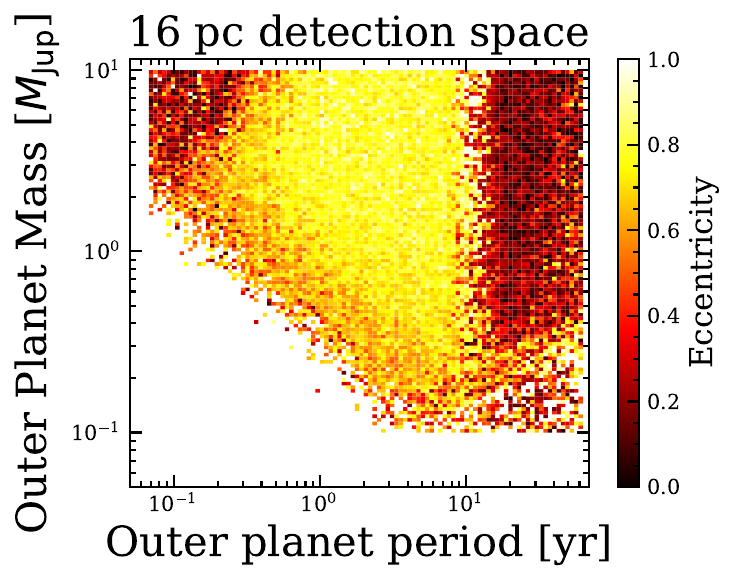}
    \caption{ Eccentricity bias from fitting a 1-planet model to data from a 2-planet system where the planets are equal in mass and are librating in dynamical resonance, for the DR5 baseline. }
    \label{fig:real_resonance_e_bias}
\end{figure}

To get the planets in dynamical resonance, we need to make sure that their resonant angle librates, where the resonant angle is defined as
\begin{equation}
    \psi = (j+k)\lambda_{2} - j \lambda_1 - k\varpi_1,
\end{equation}
where $j$ and $k$ are integers representing a $j+k:j$ resonance, $i=1$ and $i=2$ represent the inner and outer planet respectively, $\lambda_i$ is the mean longitude of planet $i$, and $\varpi_1 = \Omega_1 + \omega_1$ is the longitude of the pericentre of the inner planet, \citep{murray_solar_1999}. We use the migration prescription in ReboundX \citep{tamayo_reboundx_2020} to get the planets to reach this resonance, in the same way that we simulated GJ 876 in Section \ref{sec:2planet_example_876}, but generalising to suit any desired final orbital configuration. To generalise the approach, we initialise the outer planet at 5 times the desired semi-major axis, and set the initial period ratio with the inner planet to 2.2:1. We then set an exponential damping (proxy for migration) to the outer planet, with $\dot{a}_2/a_2 = -5 \times 10^{-5} \text{yr}^{-1}$, as well as a damping prescription for the outer planet eccentricity $\dot{e}_2/e_2 = -100 |\dot{a}_2/a_2|$ as in \citet{lee_dynamics_2002}. We include the eccentricity damping term to avoid any instabilities as the planets migrate into resonance. We scale these timescales appropriately with planet to planet mass ratio and semi major axes, to guarantee that they reach a stable resonant configuration within a reasonable integration time. We then check that the system has reached the desired orbital periods and sample Gaia data as previously described. We present the resulting eccentricity bias in Fig. \ref{fig:real_resonance_e_bias} and find that the upper left edge has indeed receded. Hence, planets in dynamical resonance will cause a bias. The remaining unaffected region escapes the bias effect due to the large libration amplitudes and important planet-planet interactions due to the large masses.

\begin{figure}
	\includegraphics[width=\columnwidth]{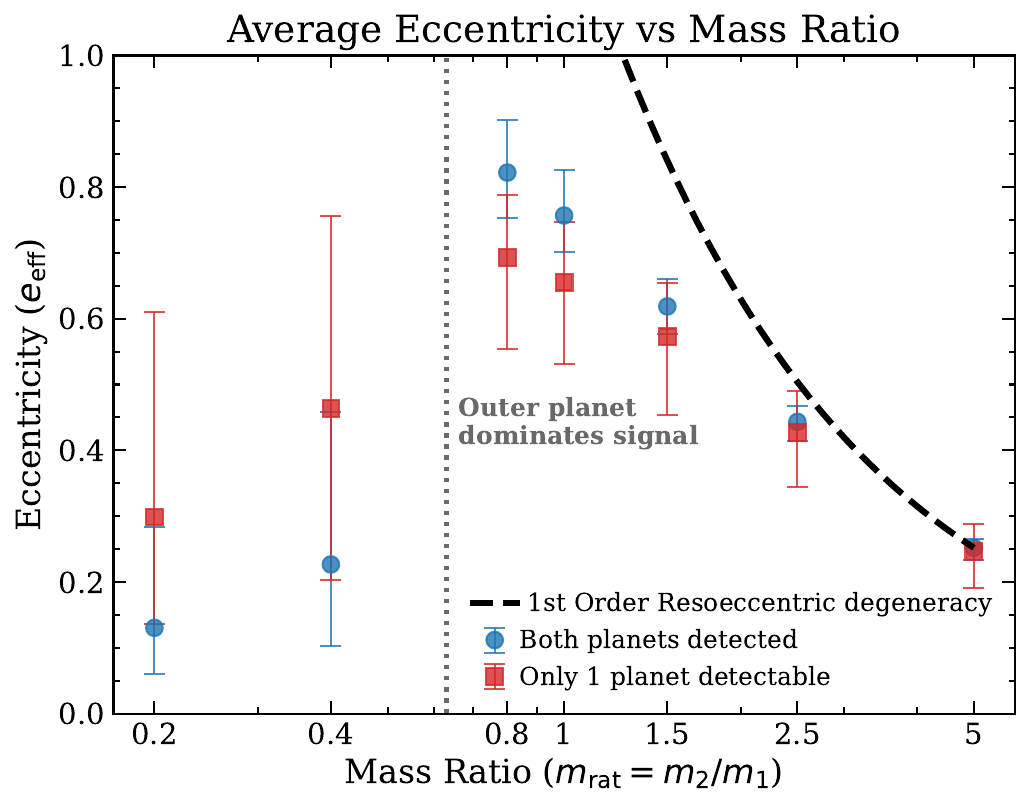}
    \caption{Median 1-planet effective eccentricity from simulated systems, with 1 standard deviation error bars, for the DR5 baseline. Taking the data for planets in dynamical resonance in Fig. \ref{fig:real_resonance_e_bias} (and similar, for each planet mass ratio) between periods of 0.4 and 8.5 years, we separate out systems where only 1 planet is statistically detectable (in red) from systems where both planets are statistically distinguishable (in blue). We also superimpose the effective eccentricity formula from \citet{yahalomi_astrometric_2025}}
    \label{fig:e_eff_vs_mass_ratio}
\end{figure}

\begin{figure}
	\includegraphics[width=\columnwidth]{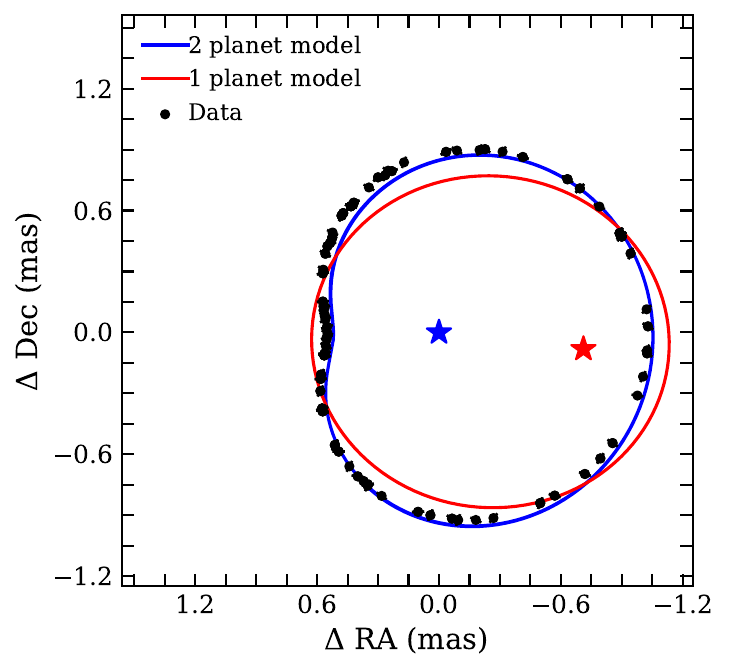}
    \includegraphics[width=\columnwidth]{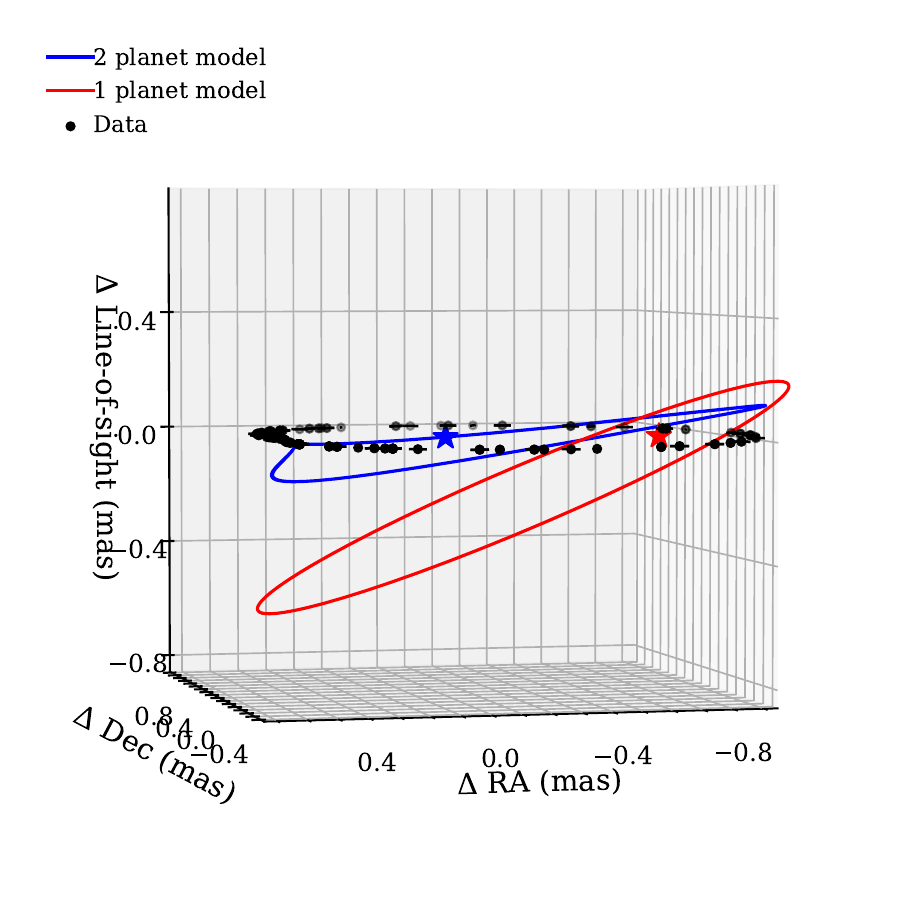}
    \caption{Astrometric reflex motion of a sample resonant 2-planet system, demonstrating the eccentricity, mass and inclination biases when fitting a single planet model to a 2-planet signal. The 2-planet fit is also represented. The original system is face-on, with a 2 $M_{\text{Jup}}$ inner planet at 2 au, near 2:1 commensurability with a 4 $M_{\text{Jup}}$ outer planet, with both planets on circular and face-on orbits. Stars represent each model's barycentre. The top plot represents the face-on line of sight view, whereas in the bottom plot it is easier to visualise the inclination, eccentricity and mass biases.}
    \label{fig:bad_1pl_fitting}
\end{figure}

\begin{figure}
	\includegraphics[width=\columnwidth]{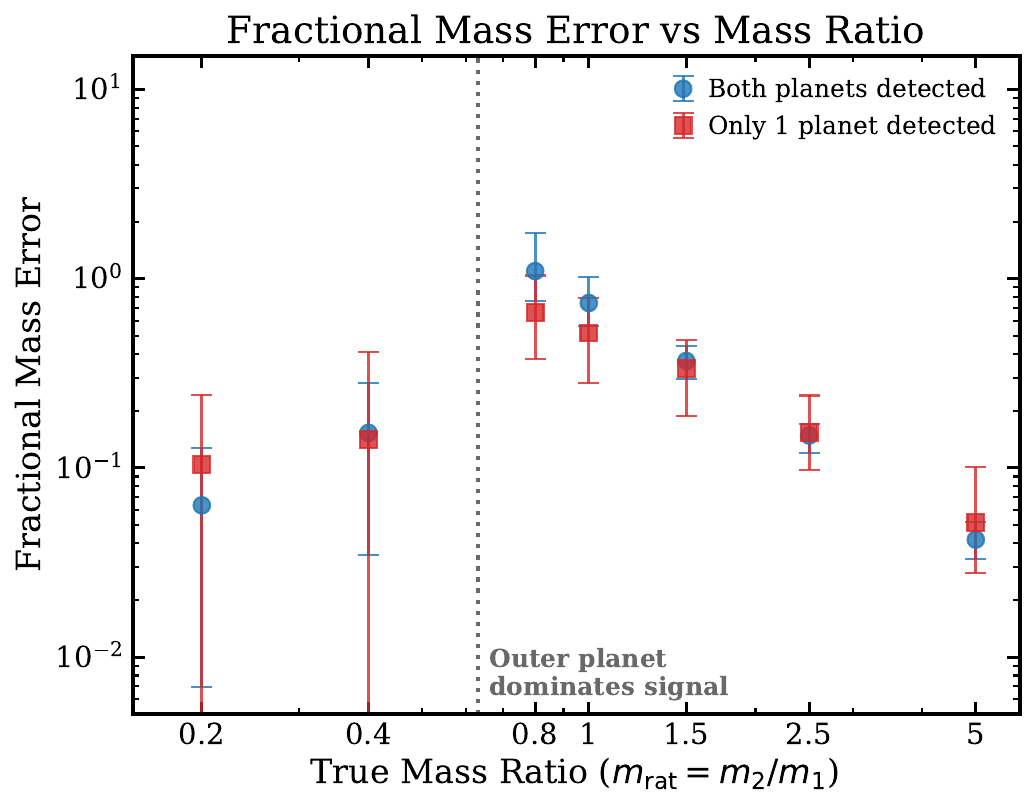}
    \caption{Median 1-planet mass fractional error from simulated systems, with 1 standard deviation error bars, for the DR5 baseline. Taking the data for planets in dynamical resonance in Fig. \ref{fig:real_resonance_e_bias} (and similar, for each planet mass ratio) between periods of 0.4 and 8.5 years, we separate out systems where only 1 planet is statistically detectable (in red) from systems where both planets are statistically distinguishable (in blue).}
    \label{fig:q_frac_err_vs_mass_ratio_filtered_dynamical}
\end{figure}

\begin{figure}
	\includegraphics[width=\columnwidth]{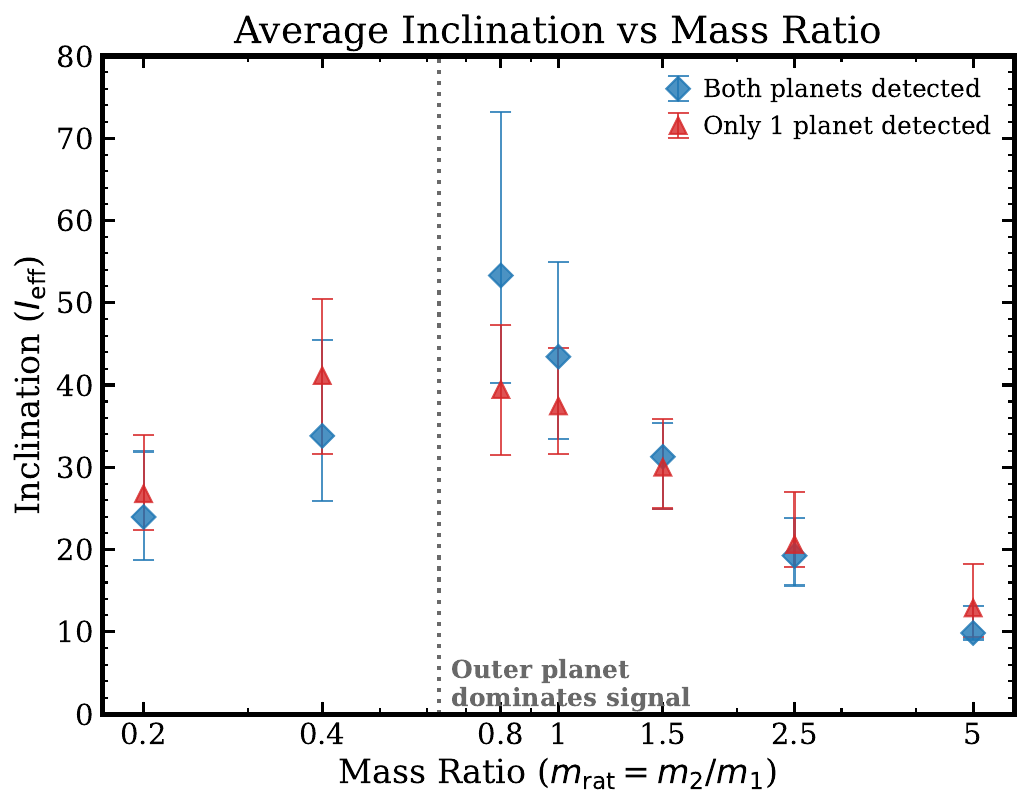}
    \caption{Median 1-planet effective inclination from simulated systems, with 1 standard deviation error bars, for the DR5 baseline. Taking the data for planets near 2:1 commensurability in the leftmost panel of Fig. \ref{fig:money_plot} (and similar, for each planet mass ratio) between periods of 2 and 8.5 years, we separate out systems where only 1 planet is statistically detectable (in red) from systems where both planets are statistically distinguishable (in blue). }
    \label{fig:i_eff_vs_mass_ratio_filtered}
\end{figure}

\subsection{Contaminated single planet signals}

We have seen that fitting a one planet signal to two planets in 2:1 resonance leads to detection biases in eccentricity and mass for identical planet masses. In this section, we aim to investigate this effect for different planet mass ratios, and quantify to what extent our Bayesian selection criteria can negate this bias.

In Fig. \ref{fig:real_resonance_e_bias}, we found the eccentricity bias region to be between 0.4 and 8.5 yrs for the outer planet, in the case of true dynamical resonances. We hence sample all the systems within this limit, and find the median effective eccentricity, plus the 1 sigma upper and lower spread. We repeat this sampling for all mass ratios, and present the results in Fig. \ref{fig:e_eff_vs_mass_ratio}. We find that the effective eccentricity peaks when the two planets produce identical astrometric signal magnitudes. We find that the biasing effect in eccentricity (and by proxy in mass) is the strongest for a mass ratio of 0.8, and more minor, although not insignificant, for the extreme mass ratio cases. We find the peak degeneracy mass ratio by considering the S/N of the resonant planets for different mass ratios:

\begin{equation} \label{eq:equal_SN}
    \frac{(S/N)_2}{(S/N)_1} = \frac{m_{2}}{m_{1}}\frac{a_{2}}{a_{1}} = \frac{m_{2}}{m_{1}} \times 2^{2/3} \approx 1.59 \times \frac{m_{2}}{m_{1}}
\end{equation}

This ratio is 1 for $\frac{m_{2}}{m_{1}} =  2^{-2/3}$, explaining why the mass ratio of 0.8 shows the strongest biases. When the planets have similar signal strengths, the amplitude of the stellar motion due to each individual planet is identical, meaning that the orbit is at its highest deviation from a Keplerian model. In Fig. \ref{fig:bad_1pl_fitting} we demonstrate this deviation by presenting the orbital motion of the star due to planets of similar signal strength, in resonance. We also show the 1 planet and 2 planet best fits. Due to the nature of resonant configurations, over several orbits the signal will repeat and always trace the same shape.

We have further split our sample into two categories: systems where we can affirm the presence of two planets and systems where only one planet is statistically detectable. For the former, the eccentricity bias is not an issue as long as the signal is correctly fit with a two planet model. In this case the measured eccentricities of both planets are close to zero. In the case of the latter, where we cannot robustly infer a second planet, there is no way to tell if the detected planet really has a large eccentricity or is simply in 2:1 resonance with an inner companion. The effective eccentricity will create a bias in the demographics. For these systems, we also find a large effective eccentricity, although smaller than in the case where both planets are detectable, generalising the results discussed from Fig. \ref{fig:money_plot}. This effect, similar to a degeneracy, is different to the resoeccentric degeneracy discussed in \citet{yahalomi_astrometric_2025}. This degeneracy is due to the weak undetectable signal from an inner companion, whereas the resoeccentric degeneracy concerns systems where both planets should in theory be clearly detectable. We find that the effective eccentricity from the resoeccentric degeneracy is only a good fit for the data at larger planet mass ratios. Mapping out these degeneracies will be the subject of future work.

In Fig. \ref{fig:q_frac_err_vs_mass_ratio_filtered_dynamical}, we find that there is also a systematic mass bias. However, this mass bias cannot be recovered by simply considering the resoeccentric degeneracy, but is minor in the high mass ratio space where the degeneracy is important. As a reference, when the outer planet mass is less than twice that of the inner planet, the mass bias is consistently greater than 10\%, and can reach up to over 100\%, as found previously. At similar planet mass ratios, this bias is slightly less important when only a single planet is statistically detectable, the same as for the eccentricity.

In Fig. \ref{fig:i_eff_vs_mass_ratio_filtered}, we plot the effective single planet inclination. For reference, the systems are all face-on, so the real inclination is 0. There is a systematic minor overestimate of the inclination (< 10 deg), due to the prior space (in reality we expect very few systems to be exactly or very near face-on due to geometric considerations). Again, as the planet masses approach the equal S/N criteria defined in Eq. \ref{eq:equal_SN}, the inclination bias rises well above the systematic limit and increases as the planet mass ratio decreases. A highly eccentric orbit has a large aspect ratio (i.e. semi-major to semi-minor axis ratio), but these 2-planet signals do not present enough elongation along the effective major axis. Hence the fit adds stretching of the aspect ratio through an effective inclination to balance out the overshooting due to the large eccentricities. The eccentricity bias is more sensitive to the time series measurements and hence is more pronounced, but the spatial extent of the orbit must also be considered here, through the inclination. This might provide a route to quantifying the impact of this degeneracy at the population level, as one would naturally expect the inclinations of real exoplanet systems to be isotropically distributed on the sky or, alternatively, to be linked to the spin axis of the star.

\begin{figure}
	\includegraphics[width=\columnwidth]{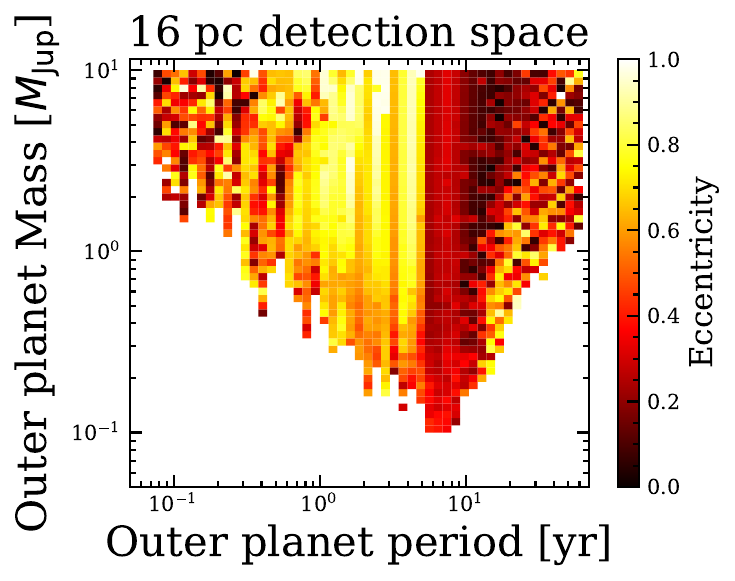}
    \caption{ Eccentricity bias from fitting a 1-planet model to data from a 2-planet system where the planets are equal in mass and near 2:1 commensurability, for the DR4 baseline. }
    \label{fig:DR4_money_2t1}
\end{figure}

The results shown in Figs. \ref{fig:e_eff_vs_mass_ratio}, \ref{fig:q_frac_err_vs_mass_ratio_filtered_dynamical}, and \ref{fig:i_eff_vs_mass_ratio_filtered} raise an important question: when only a single planet can be statistically detected, is it possible or not to determine whether or not its eccentricity is real or is the effect of a second undetectable planet?

The answer to this question is far from obvious, and RV follow-ups might help break the degeneracy. However, these follow-ups will require a lot of observational effort. We present here an idea that could help identify potential targets that demonstrate hints of eccentricity-mass-inclination bias. We notice in Fig. \ref{fig:money_plot} that the eccentricity bias drops when the outer planet period is longer than the observing baseline, in this case, the 10 years of DR5. In Fig. \ref{fig:DR4_money_2t1} we draw the same plot with the eccentricity bias, but this time only considering the time series available in DR4. We find that this outer edge lies at an outer-planet period of approximately 5 years. Then, comparing the eccentricities for all systems with periods between 5 and 10 years, we expect the effective eccentricity from DR5 to be large and hence biased, whereas the effective eccentricity from the DR4 data would be close to 0. This idea relies on the caveat that beyond periods of 5 and 10 years we are still able to accurately detect and characterise planets with DR4 and DR5 respectively. This is possible, as seen in the discussion of single planet fits but will induce large uncertainties in the retrieved orbital parameters. 

\begin{figure}
	\includegraphics[width=\columnwidth]{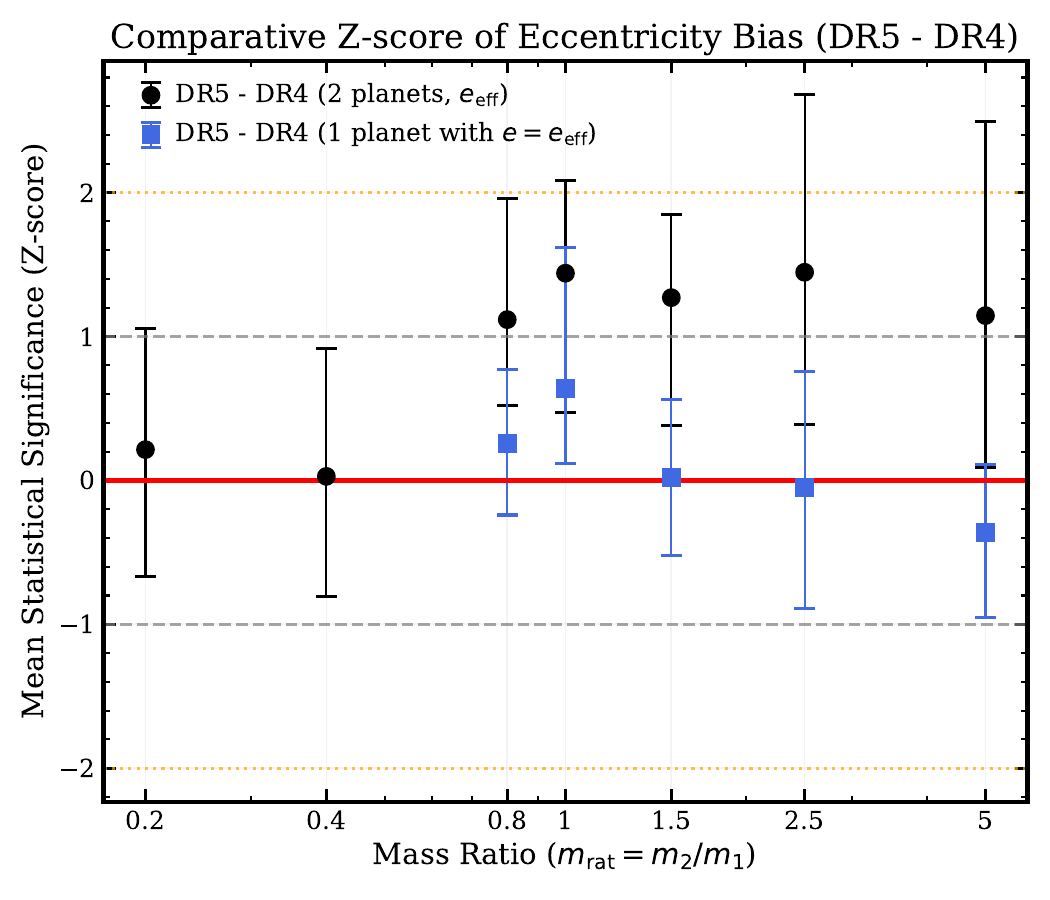}
    \caption{Mean $Z$-score of the difference between the eccentricities recovered from one-planet fits over the DR5 and DR4 baselines (DR5 minus DR4), shown as a function of planet mass ratio. Black circles represent two-planet input signals, while blue squares represent matched synthetic one-planet signals. For each synthetic one-planet signal, the injected eccentricity is set to the effective eccentricity recovered from the DR5 one-planet fit to the corresponding two-planet system. Markers show the mean $Z$-score and error bars span the 16th--84th percentiles of the individual-system distribution. The parent two-planet systems are restricted to those for which only one planet is detectable over both baselines and for which the original outer planet has an orbital period of $5$--$10\,\mathrm{yr}$, corresponding to the assumed DR4 and DR5 baselines. The blue points additionally require successful matching and detection in both synthetic one-planet datasets.}
    \label{fig:comparative_mean_zscore_corrected_pipelines}
\end{figure}

To explore the idea, we run another series of tests. We note the effective eccentricity for the planets in the DR5 grid (Fig. \ref{fig:money_plot}), and generate a series of single planet signals with the same orbital parameters as the largest S/N planet from the grid study, but with the eccentricity set to the effective eccentricity from the 1 planet fit to the 2 planet signal. We fit these signals with single planet models for a DR4 and DR5 time series. We hence have 4 setups: two 1-planet fits (DR4 and DR5) to a circular 2-planet signal, and two 1-planet fits (DR4 and DR5) to an eccentric 1-planet signal. We show our results in Fig. \ref{fig:comparative_mean_zscore_corrected_pipelines}, as a Z score for each individual system, where the mean value is shown with the $1-\sigma$ statistical spread in the data:

\begin{equation}
    Z = \frac{e_{\text{DR5}}-e_{\text{DR4}}}{\sqrt{\sigma_{e_{\text{DR5}}}^2+\sigma_{e_{\text{DR4}}}^2}}
\end{equation}

We consider the difference between the DR4 and DR5 eccentricities, $e_{\text{DR4}}$ and $e_{\text{DR5}}$, respectively. If the difference itself is larger than the uncertainty on the difference, then we can confirm a hint for a hidden second planet. On average, in the case of the 2-planet fits, the difference in the eccentricity is between 1 and 2 times larger than the value of the uncertainty, with a consistent systematic bias towards larger DR5 eccentricities. We conclude that there can be moderate statistical evidence for a second planet. When we apply the same analysis to the 1 planet signals, we find most systems have an uncertainty larger than the eccentricity difference, so either both eccentricities are small or the solver finds the same values for DR4 and DR5, which is not always the case for the 2-planet fits. 

We conclude that this method could tentatively work for certain specific systems, and is worth exploring further. However, we discuss a few nuances. The 1 planet systems were drawn without consideration of the mass bias, so including this effect would lead to better planet characterisations, reducing eccentricity uncertainties. Additionally, in the context of DR4, we can imagine degrading this search to periods between 3 and 5 years, for example, however, the uncertainties in the 3 year dataset would massively hinder the ability to obtain large Z-scores. Applications of this method to real data will hence have to wait until DR5.

\subsection{Do resonances hinder multi-planet detection capabilities?} 

\begin{figure*}
	\includegraphics[width=\textwidth]{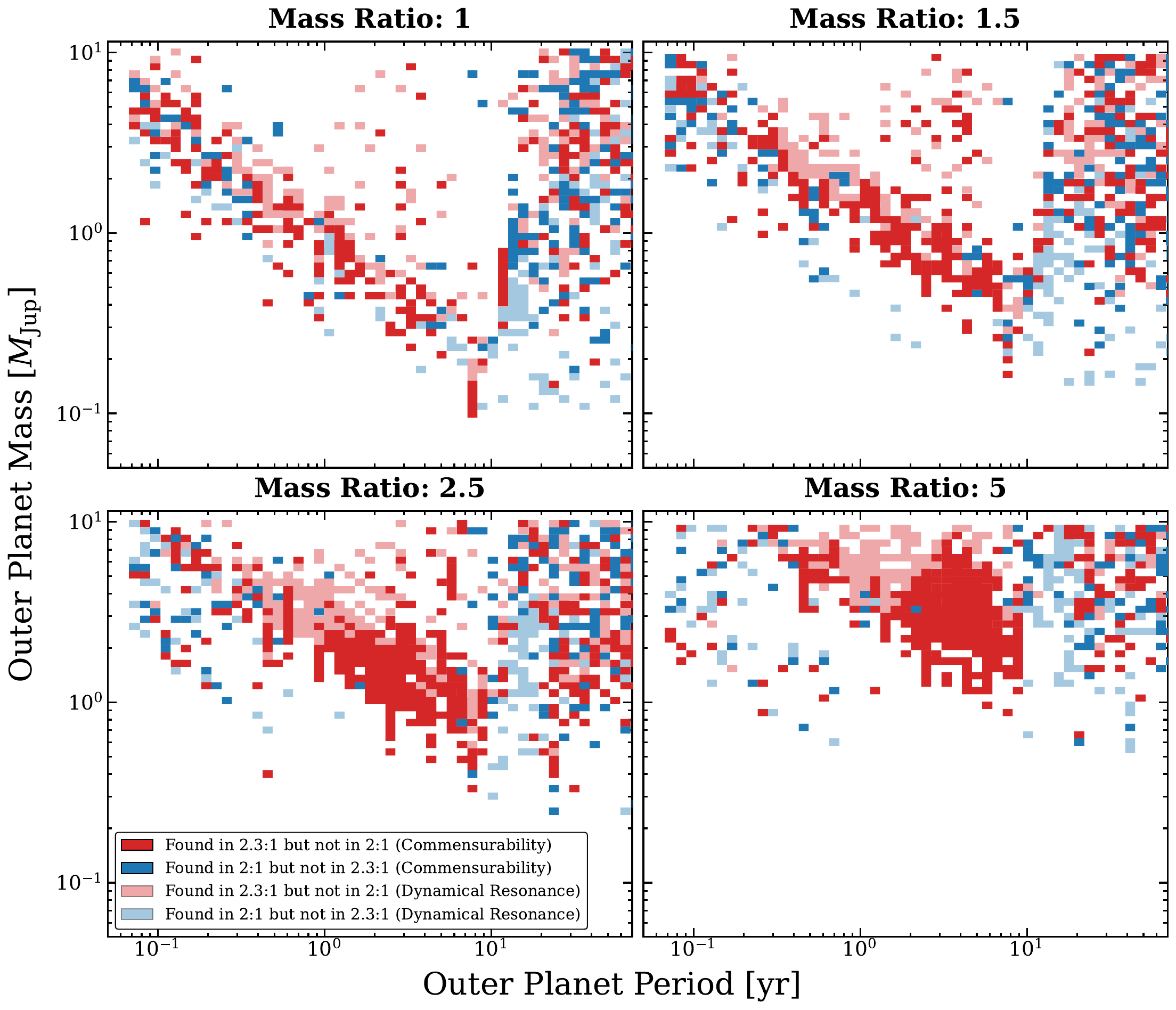}
    \caption{Comparison of the detectability of the second planet near the 2.3:1 and 2:1 period commensurabilities, shown as a function of outer-planet orbital period and mass for four simulated mass ratios, $m_2/m_1$. Opaque cells compare systems set up at commensurability: red denotes detection at 2.3:1 but not at 2:1, while blue denotes the reverse. Translucent cells show the corresponding comparison after replacing the commensurable 2:1 systems with matched, dynamically evolved 2:1-resonant systems. A detection requires $\log_{10}{BF}_{21}>2$, indicating a preference for the two-planet model. White regions indicate the same detection outcome for both configurations or the absence of a matched comparison.} 
    \label{fig:resonance_isworse}
\end{figure*}

\begin{figure*}
	\includegraphics[width=\textwidth]{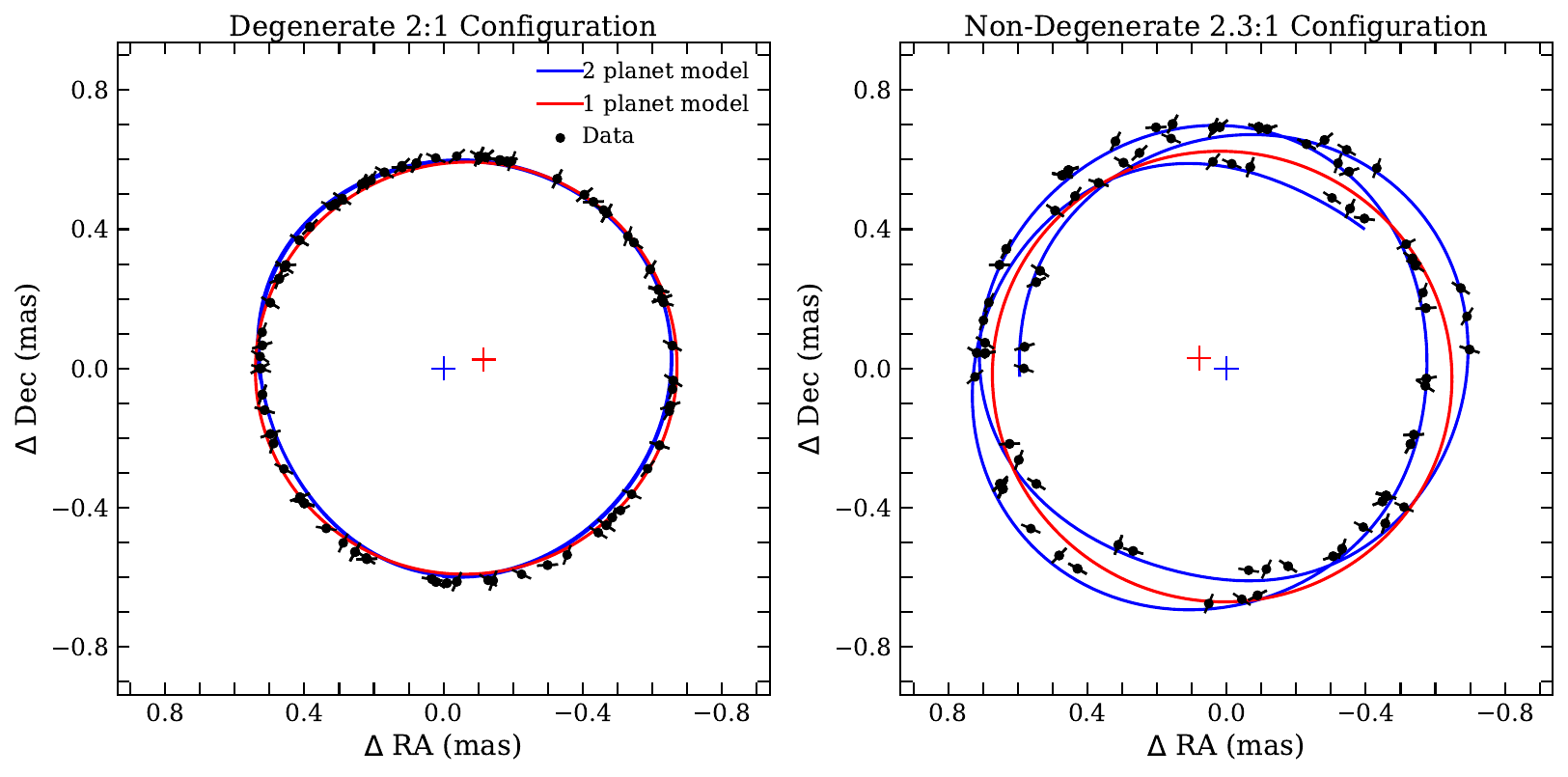}
    \caption{Astrometric reflex motion of a sample system where the resoeccentric degeneracy applies. 1- and 2-planet fits are shown; red and blue crosses indicate their respective barycentres. Left: data and fits for a system near 2:1 commensurability, with a 0.8 $M_\text{Jup}$ inner planet at 1.5 au and a 4 $M_\text{Jup}$ outer planet. There is no statistical evidence to favour the 1-planet versus the 2-planet model. Right: Same as in the left panel, but with the outer planet near 2.3:1 commensurability. The 2-planet model is statistically preferred.}
    
    \label{fig:res_degen}
\end{figure*}

For large separation planets (P>1 yr), Gaia will observe fewer than 10 orbits. We want to know whether or not having the two planets in 2:1 resonance will affect the detectability of the multi-planet system. In Fig. \ref{fig:resonance_isworse}, we compare the relative detectability of systems in and out of resonance (with period ratios of 2:1 and 2.3:1 respectively), for different planet mass ratios. In blue we show systems that are detectable when the period ratio is 2:1, but not when it is 2.3:1. In red we show the opposite, i.e. systems that are detectable in 2.3:1 period ratio but not in 2:1. To include the effects of dynamical resonance, we also map in a lighter hue the same colours, but comparing the 2.3:1 commensurability setup to the setup described in Section \ref{sec:librations}, in dynamical resonance.

We notice the greatest difference in the case of a large mass ratio ($m_{\text{out}}/m_{\text{in}} = 5$). There are more systems that are detectable only out of resonance than systems that are exclusively detected in resonance. These undetectable resonances are more prevalent in the red area of the bottom right panel of Fig. \ref{fig:resonance_isworse}, extending from 0.3 to 10 years, and from 1 to 10 $M_\text{Jup}$. The larger the outer planet period, the lower the minimal outer planet mass necessary for the resonance to be undetectable, in other words, at larger periods the degeneracy covers a larger range of planet masses.

For mass ratios of unity (top left panel) or below, there is no visible evidence for a paucity of resonant system detections. As planet mass ratio increases, the plots present more undetectable resonant systems. They populate a diagonal, which gets thicker at larger planet mass ratios. This diagonal is populated by systems where only a single planet is statistically detectable. We hence conclude, as the outer planet gets comparatively more massive than the inner planet, that both planets become increasingly difficult to detect together. We can no longer statistically disambiguate the two planet signal from the one planet signal.

Additionally, we find that systems in dynamical resonance push this degeneracy further to shorter periods compared to planets simply in a 2:1 commensurability.

To explain this degeneracy, we invoke geometric and mathematical arguments. In Fig. \ref{fig:res_degen}, we show the fitting process for a system within the degenerate parameter space. Only 3 outer planet orbits are observed and we compare the shape of the orbit for large planet mass ratios in and out of resonance. When the system is in resonance, the signal looks similar to an ellipse (we will show later that to first order it is), with a small deviation due to the inner, barely distinguishable planet. Since the planets align regularly due to the resonance, this deviation is consistently located at the same orbital longitude, and hence is not statistically significant. Contrary to Fig. \ref{fig:bad_1pl_fitting} where the planet masses are closer in value, the large relative signal strength in this case (the outer planet has a S/N 8 times larger than that of the inner planet) leads to an inconclusive 2-planet fit. When examining Bayes factors, we find that the evidences are similar between the two models, so that neither model is statistically preferred, indicative of a true degeneracy. On the other hand, when the systems are out of resonance, over a few orbits the longitude of conjunction greatly varies along the orbit, scattering the small deviation to affect the entire orbit. The resulting orbital signal is hence more difficult to explain with a 1-planet model only. We then tend to prefer the 2-planet model, as seen in the right panel of Fig. \ref{fig:res_degen}. 

Mathematically, we find that the affected region of the parameter space falls within the bound of resoeccentric degeneracy \citep[][]{yahalomi_astrometric_2025}. Following \citet{yahalomi_astrometric_2025}, by expanding Eq. \ref{eq:kepl_orbit} to first order in $e$ and entering the result in Eq. \ref{eq:ra_decl}, we find that for a single planet

\begin{equation}
\begin{aligned}
\Delta\alpha^*(t) \approx {} & \frac{a_\star}{d} [B' \cos M(t)  + G' \sin M(t) ] \\
& + \frac{1}{2} e \frac{a_\star}{d} [B' \cos 2M(t)  + G' \sin 2M(t) ] \\
& - \frac{3}{2} e \frac{a_\star}{d} B'.
\end{aligned}
\end{equation}
When considering a 2 planet system with no eccentricities, Eq. \ref{eq:XY_2pl} simplifies to 
\begin{equation}
\begin{aligned}
\Delta\alpha_{\mathrm{tot}}^*(t) \approx {} & \frac{a_{\star,2}}{d}[B' \cos M (t) + G' \sin M(t) ] \\
& + \frac{a_{\star,1}}{d}[B' \cos 2M(t)  + G' \sin 2M(t) ],
\end{aligned}
\end{equation}
assuming identical Thiele-Innes constants for both planets, where $M(t)$ is now defined as the outer planet's mean anomaly. By comparing each frequency component (and similarly in declination), we can establish the following relationship between the effective eccentricity of the single eccentric planet and the planet mass ratio:
\begin{equation}
    e_{\text{eff}} = 2^{1/3} \frac{m_{\text{in}}}{m_{\text{out}}}
\end{equation}
such that, for example, in the case where $m_{\text{out}}/m_{\text{in}} = 5$, $e_{\text{eff}} = 0.25$, which is what we observe from our simulations in Fig. \ref{fig:e_eff_vs_mass_ratio}. In this case, for high mass ratios, we find that the resoeccentric degeneracy is an appropriate description of the bias. This discussion, however, is incomplete; in the case with equal mass planets, $e_{\text{eff}} > 1$, which is a result of the first order approximation in $e$. Other approximations such as identical Thiele-Innes coefficients, circular resonant planets, and neglecting planet-planet interactions contribute to weakening the applicability of the resoeccentric degeneracy. From Fig. \ref{fig:e_eff_vs_mass_ratio}, we already find significant deviations in effective eccentricity below a mass ratio of 2.5. Investigating the relevant parameter space and validity of approximations in more detail is a task for future work. The most probable explanation for the non-detections in the equal-mass systems would be that in the case of 2 low S/N signals, only the strongest one is statistically detectable.

When looking at resonances, we can also consider the case of low-period (P < 1 yr) planetary systems, with a more massive outer planet. A good case of this would be the resonant planets b and c from the well-studied GJ 876 system ($m_{\text{b}}/m_{\text{c}} = 3.2$), as already discussed in Section \ref{sec:2planet_example_876}.
In this system, planet c is the one with the lowest signal strength. We found that it is effectively detectable in DR4, but further comparisons with DR3 and DR5 baselines are necessary to quantify when the planet is detectable, and if not, to find out how it affects the retrieved parameters of planet b. We also want to investigate to what extent librations will affect the retrieved planetary parameters.

We simulate the system again, in the same way as described in Section \ref{sec:2planet_example_876}. We now fit the resulting signal with a 1-planet and 2-planet models, for the observation baselines of DR3, DR4 and DR5. Our results in eccentricity are presented in Table \ref{tab:gj876_e}.

We chose this specific system because we can compare our results with those of \citet{holl_gaia_2023}, where the currently unreleased DR3 astrometry time series are fit for a handful of systems, including GJ876. They fit a single planet model to 36 months of data and find $e_{\text{b}} = 0.16 \pm 0.15$. Although more precise measurements of eccentricity ($e_{\text{b}} \sim 0.037$ from \citealt{nelson_empirically_2016}) fall within this range, it could be that the result is  biased. This bias can have two distinct origins: it could either be due to the influence of unmeasured planet c, or it could be a consequence of the fit to the uneven sampling. In \citet{holl_gaia_2023}, the authors account for the latter by truncating the eccentricity prior, thereby also mitigating the effect of the resonant planet c. We do not truncate our prior, explaining why we find a larger value of eccentricity when fitting a 1 planet model to the DR3 data, consistent with the resoeccentric degeneracy. We find similar results for DR4. By truncating the eccentricity prior we mitigate the biasing in planetary parameters, but also prevent finding traces of additional planets. In DR5 however, the single planet fit does not show traces of a degeneracy. We believe that the longer baseline has allowed Gaia to observe resonant librations and hence break the degeneracy. Given the apsidal precession rate of roughly 41 deg/yr \citep[][]{lee_dynamics_2002,laughlin_gj_2005}, over DR4 we will observe 57\% of an apsidal precession cycle and over DR5 we will observe 114\% of the duration of the precession cycle. In our simulated setup, we measure the precession rate to be lower, i.e. 14 deg yr$^{-1}$ with a libration amplitude of < 4 degrees for both planets. Effectively, in our simulations we measure less than 40\% of a precession cycle for DR5. Further work is needed to establish the precession rate thresholds at which the degeneracy may break, but we have already shown that some Gaia data might fall within this regime.

In DR3, we cannot statistically distinguish planet c, but in DR4 and DR5 we are able to statistically find a second planet, thereby reducing the outer planet eccentricity to near 0, and finding an inner planet eccentricity consistent with current RV observations.

\begin{table}
\centering
\caption{Comparing the eccentricities for the 1 and 2 planet fits of the mock GJ876 data for DR3, DR4 and DR5. The value in the reference e column for the single planet model corresponds to the predicted effective eccentricity from \citet{yahalomi_astrometric_2025}. The measured values are from \citet{nelson_empirically_2016}. }
\label{tab:gj876_e}
\begin{tabular}{l c c c c}
\hline
 & \textbf{Reference e} & \textbf{DR3} & \textbf{DR4} & \textbf{DR5} \\
\hline
 single planet & $e_{\text{eff}}= 0.39$ & $0.464^{+0.079}_{-0.085}$ & $0.360^{+0.061}_{-0.066}$ & $0.058^{+0.043}_{-0.036}$ \\
\hline
 inner planet c & $e_{\text{meas}}=0.25$ & $0.267^{+0.064}_{-0.045}$ & $0.122^{+0.137}_{-0.086}$ & $0.138^{+0.128}_{-0.090}$ \\
 outer planet b & $e_{\text{meas}}=0.037$ & $0.002^{+0.001}_{-0.001}$ & $0.017^{+0.047}_{-0.015}$ & $0.015^{+0.028}_{-0.012}$ \\
\hline
$\log_{10} BF_{21}$ & & -2.2 & 6.3 & 26.6 \\
\hline

\end{tabular}
\end{table}



\section{Discussion and further work} \label{sec:discussion}

From the results presented in this study, we see tentative evidence that the planet-planet interactions and apsidal precession resulting from strong secular interactions and enhanced by the resonance play an additional role in the data, and cannot be completely neglected. For systems like GJ 876, we have seen that the baseline of DR5 covers more than one precession cycle for the real system. We hence motivate the need for more detailed modelling of such systems. Including an N-body fit or a proxy for planet-planet interactions might allow us to statistically infer the presence of these planet-planet interactions. For short period planets, the pericentre longitude change might lead to along-scan positional deviations on angular scales greater than the Gaia astrometric precision over 5 to 10 years. Additionally, even if the libration amplitude is on the order of degrees, in extreme cases like GJ 876 it is known that its libration period is on the order of a year \citep[][]{correia_harps_2010} and we can still expect potentially measurable delays in pericentre passage due to this libration, which can cause significant residuals for a two Keplerian model. These variations can motivate the need for an analogous photodynamical modelling technique used in transits \citep[e.g.][]{Carter2012}, which could lead to identifying librations or hints of librations and contribute to identifying true resonant systems amongst the giant planets found by Gaia. However, we want to highlight a few caveats to this method. Unlike transit observations, Gaia samples sparsely and on a scan-law-determined cadence, so the modulation must neither alias against the scanning law nor be so slow that only a small fraction of a cycle falls within the 5 to 10 yr baseline. Additionally, GJ 876 is an outlier regarding resonance detectability. For most resonant systems, this analysis would not be fruitful, and would come at a significant computational cost. A rigorous adjustment of the prior space for both interacting and non-interacting models is hence necessary to provide meaningful Bayes factors that identify deviations. However, it is still worth exploring in future work.

The various biases identified in this work will have a significant impact on the demographics from Gaia. It is therefore important to quantify these biases. The least significant of them all, the inclination bias, can already inform us on what fraction of systems are subject to all other biases. If we notice an inclination distribution that deviates from a purely geometric expectation, then we will have a clear hint that the biases affect demographics. To further adjust the demographics, it is necessary to find a mathematical model that goes beyond the resoeccentric degeneracy and encapsulates the biases at large effective eccentricities, in both cases where the second planet is either detectable or undetectable. With this model, adjusting the giant planet occurrence rates on a population scale hence becomes possible.

Finally, we want to address various important technical points that were neglected in the detection pipeline. The modelling of the error is assumed to be random normally distributed uncorrelated noise, and hence ignores the various scan angle biases identified in Section \ref{sec:dr4_notebook}. To obtain meaningful detections from actual DR4 data would require to integrate scan-angle harmonics into the different model comparisons, as well as a jerk term on top of the linear acceleration term already implemented. These effects are already included in the detection pipeline presented in \citet{baycroft_bayesian_2026}, where the scan angle signal identified in Target 4 of Section \ref{sec:dr4_notebook} is correctly recovered. These additions will significantly reduce contamination by false positives due to unresolved binaries, or any other scan angle related signals. Additionally, DR4 orbit fits from the pipeline will also have to be compared with models fixing a circular orbit, to confirm that the planets are indeed observed on eccentric orbits as opposed to a spurious eccentricity arising from the data S/N or uneven sampling cadence. All of the previous adjustments will have to be made to the pipeline in order to apply it rigorously to real data, on top of increasing the number of live points to achieve credible and systematic detections.

\section{Summary}  \label{sec:conclusions}

We used the N-body code REBOUND to simulate astrometric data as measured by Gaia for giant multi-planet systems at au distances. We then developed a planet detection pipeline based on nested sampling to fit different planetary architectures, including a purely astrometric solution with and without a linear acceleration, a 1-planet model and a 2-planet (non-interacting Keplerian) model. Our pipeline was verified on the published time series Gaia data from the BH3 detection, where we accurately recovered its parameters. 

Using our pipeline we recover previously established 1-planet detection criteria with Gaia. We also investigated 2-planet detection criteria with a grid study, with particular attention to systems near and far from 2:1 resonance. We sought to identify each scenario where 2-planet detections are statistically impossible and how a second undetected planet would affect the retrieved orbital parameters for the other planet.

We found for single planet detections that we are able to detect shorter period planets (P < 10 yrs) when they have an astrometric S/N per transit greater than 1.01 and 0.77 for DR4 and DR5 baselines, respectively. Including a linear acceleration, we find a sharp cut-off in system detectability for planets beyond 6 au and 10 au for DR4 and DR5 respectively.

We found that Gaia is capable of detecting 2-planet systems, using synthetic GJ 876 data as an example. Investigating a grid of 2-planet systems spanning a range of planet mass and period ratios, we found that the lower S/N planet is harder to detect. In DR5, this planet is detected at S/N per astrometric point threshold between 0.5 and 2, as opposed to a range of 0.5 to 0.75 for the dominant planet. We also found that planets closer to 2:1 commensurability are harder to find, and systems with a more massive outer planet show larger deviations from the theoretical detectability power law.

When fitting a 1-planet model to data from a 2-planet system in 2:1 resonance, we found that we recover the dominant planet in the signal, with significant eccentricity, mass and inclination biases. These biases become more important as the astrometric signatures of both planets get closer ($e_{\text{bias}} = 0.8$ and 0.4 for mass ratios of 0.8 and 2.5 respectively). When the planets are closer in mass, in most cases both planets can be recovered without bias. However, when one planet becomes statistically undetectable the bias persists. When the outer planet is at least three times as massive as the inner planet, the bias tends towards the previously known resoeccentric degeneracy, where 1-planet and 2-planet solutions are degenerate. We found that placing the planets in dynamical mean-motion resonance extends the parameter space for this degeneracy down to shorter periods. Finally, using the example of GJ 876, we found that apsidal precession can break the eccentricity bias if the system is observed for a long enough baseline, i.e. that of DR5.

Overall, if giant multi-planet systems are common the Gaia catalogue might end up being plagued with single planet systems with anomalously large eccentricities. A thorough re-analysis of the DR4 data might be necessary to confirm if certain targets, especially those with large residuals, are better explained as resonant 2-planet systems. The severe mass bias (between 10\% for mass ratios of 2.5 and 100\% for mass ratios of 0.8) will have a significant impact on the inferred giant planet demographics from the Gaia catalogue. This result will bias the observational constraints used to test planet-formation models if not corrected, as we might falsely expect there to be fewer systems in 2:1 resonance from the Gaia data, compared to the real distribution of these planets. To identify the degree of contamination in the data, a follow-up idea could be to compare the population-wide distribution of inclinations (including the bias) to the distribution expected from geometry only, taking into account the harder detectability of edge-on systems. 

\section*{Acknowledgements}

 This project has received funding from the European Research Council (ERC) under the European Union’s Horizon 2020 research and innovation programme (Grant agreement No. 853022). EG has received support from an STFC PhD studentship. JEO is supported by a Royal Society University Research Fellowship. This work used the DiRAC Data Intensive service (CSD3) at the University of Cambridge, managed by the University of Cambridge University Information Services on behalf of the STFC DiRAC HPC Facility (www.dirac.ac.uk). The DiRAC component of CSD3 at Cambridge was funded by BEIS, UKRI and STFC capital funding and STFC operations grants. DiRAC is part of the UKRI Digital Research Infrastructure. This work has made use of data from the European Space Agency (ESA) mission Gaia (https://www.cosmos.esa.int/gaia), processed by the Gaia Data Processing and Analysis Consortium (DPAC, https://www.cosmos.esa.int/web/gaia/dpac/consortium). Funding for the DPAC has been provided by national institutions, in particular the institutions participating in the Gaia Multilateral Agreement.

\section*{Data Availability}

Data from our numerical models are available upon reasonable request to the corresponding author.



\bibliographystyle{mnras}
\bibliography{example} 


\bsp	
\label{lastpage}
\end{document}